\documentclass[acmlarge]{acmart} 
\setcopyright{none}

\acmJournal{IMWUT}
\acmYear{2027}
\acmVolume{0}
\acmNumber{0}
\acmArticle{0}
\acmMonth{0}

\usepackage{amsmath,amssymb,amsfonts}
\usepackage{graphicx}
\usepackage{textcomp}
\usepackage{booktabs}
\usepackage{multirow}
\usepackage{makecell}
\usepackage{pifont}
\usepackage{algorithm}
\usepackage{algpseudocode}

\newcommand{\cmark}{\ding{51}}%
\newcommand{\xmark}{\ding{55}}%

\usepackage{xcolor}

\usepackage{subcaption}
\usepackage{comment}

\begin{document}

\title{Ultrasound-Assisted Biometric Watermarking Against Face Video Deepfakes}

\author{Guodong Liu}
\orcid{0009-0004-2463-9586}
\affiliation{%
  \institution{Nanjing University}
  \city{Nanjing}
  \state{Jiangsu}
  \country{China}
}
\email{EricLiu@smail.nju.edu.cn}

\author{Lei Wang}
\authornote{Corresponding authors.}
\orcid{0000-0002-0091-0931}
\affiliation{%
  \institution{Soochow University}
  \city{Suzhou}
  \state{Jiangsu}
  \country{China}
}
\email{wanglei@suda.edu.cn}

\author{Dongxu Liu}
\orcid{0009-0008-8305-7198}
\affiliation{%
  \institution{Nanjing University}
  \city{Nanjing}
  \state{Jiangsu}
  \country{China}
}
\email{dongxuliu@smail.nju.edu.cn}

\author{Qianran Qiao}
\orcid{0009-0009-4767-5922}
\affiliation{%
  \institution{Nanjing University}
  \city{Nanjing}
  \state{Jiangsu}
  \country{China}
}
\email{522025330078@smail.nju.edu.cn}

\author{Meng Li}
\orcid{0000-0001-5764-960X}
\affiliation{%
  \institution{Nanjing University}
  \city{Nanjing}
  \state{Jiangsu}
  \country{China}
}
\email{meng@nju.edu.cn}

\author{Haipeng Dai}
\authornotemark[1]
\orcid{0000-0003-0545-8187}
\affiliation{%
  \institution{Nanjing University}
  \city{Nanjing}
  \state{Jiangsu}
  \country{China}
}
\email{haipengdai@nju.edu.cn}

\author{He Huang}
\orcid{0000-0003-2768-6607}
\affiliation{%
  \institution{Soochow University}
  \city{Suzhou}
  \state{Jiangsu}
  \country{China}
}
\email{huangh@suda.edu.cn}

\author{Guihai Chen}
\orcid{0000-0002-6934-1685}
\affiliation{%
  \institution{Nanjing University}
  \city{Nanjing}
  \state{Jiangsu}
  \country{China}
}
\email{gchen@nju.edu.cn}


\begin{abstract}
Currently, artificial intelligence-generated content (AIGC) poses a severe threat to facial video deepfakes, and many AIGC companies employ video watermarking to mitigate unauthorized fake video usage, but its effectiveness is limited due to watermark tampering and private-model forgery.
Additionally, some schemes actively add watermarks to user videos. However, current approaches focus primarily on the quality of watermark embedding in offline scenarios, and no existing work offers a solution for protecting human facial videos in real-time online applications.
To fill this gap, we propose, to the best of our knowledge, a novel biometric video watermarking method, \textbf{UltraGuard}, for the first time. 
It uses the smartphone’s built-in speaker and microphone to emit ultrasonic signals and capture biometric features, which are then fused with spatio-temporal dynamic features derived from the video content.
The fused features are encoded as watermarks and embedded into videos in real time, enabling dynamic, content-aware watermarking throughout transmission and propagation.
Extensive experimental validation demonstrates that UltraGuard can effectively defend against malicious attacks such as forgery and editing, both at the source and during propagation, proving the system's efficacy. 
In the future, it can be deployed on users' mobile devices to provide real-time protection for users' facial videos, thereby preventing security and privacy breaches caused by various malicious attacks.
\end{abstract}



\maketitle

\section{Introduction}
 
With the widespread adoption of high-speed mobile networks, online video communication, such as video conferencing and real-time video calls, has become integral to modern information exchange.
In many practical scenarios, these interactions involve legal archiving and business decision-making, which makes complete and trustworthy video records increasingly important. 
At the same time, the security and authenticity of video content become more difficult to guarantee \cite{zhang2022deepfake}. 
To counter highly deceptive Deepfake and malicious tampering techniques, existing defenses increasingly shift from traditional passive detection to data-driven deep learning methods \cite{passos2024review}. 
However, these approaches often show limited generalization when they face unseen cross-domain attacks or complex transmission distortions \cite{khan2023deepfake}.
Furthermore, we note that with the recent advancement of AIGC, representative models such as Sora \cite{openai2024sora} and Seedance \cite{gao2025seedance} can generate forged facial videos that are increasingly difficult to distinguish from real ones.
The latest face-swapping algorithm \cite{flynn2026edityourself, zhang2026uniswap} focuses specifically on modifying target faces, which will make forgery detection increasingly challenging in the future, posing a significant threat to user privacy and security.
Against this backdrop, digital video watermarking emerges as a promising direction for proactive provenance tracking and content authentication \cite{zhang2011spread, lai2025enhancing}.
However, the mainstream solution adopted by current large model vendors is to embed watermark information into forged videos. However, this approach is easily removed by attackers or circumvented by privately trained generative models. A better facial anti-spoofing solution is to proactively embed watermark information into all genuine human face videos of users, such that any video with an incorrect watermark or without a watermark is regarded as an attacker-forged video.

Furthermore, a closer look at prior work reveals fundamental security limitations in existing video watermarking schemes. Most of them are designed as static watermarking solutions for offline settings \cite{zhang2011spread, fernandez2024video, hu2025videomark}, treating the watermark as a post hoc tag rather than a source-authenticating signal at capture time. As a result, they cannot verify whether a video truly originates from a live, legitimate source.
This weakness leaves them vulnerable to replay attacks, allowing pre-recorded forged videos to obtain valid watermarks and be accepted as trusted content \cite{aberna2024digital,asikuzzaman2017overview}.
Moreover, because most video watermark extractors are publicly accessible, attackers can reverse-engineer the extraction model, infer static watermark features, and reconstruct them for injection into tampered videos. This can mislead the verifier into returning a valid decision, thereby enabling highly stealthy forgeries.

\begin{figure}[t]
  \centering
  \includegraphics[width=0.8\linewidth]{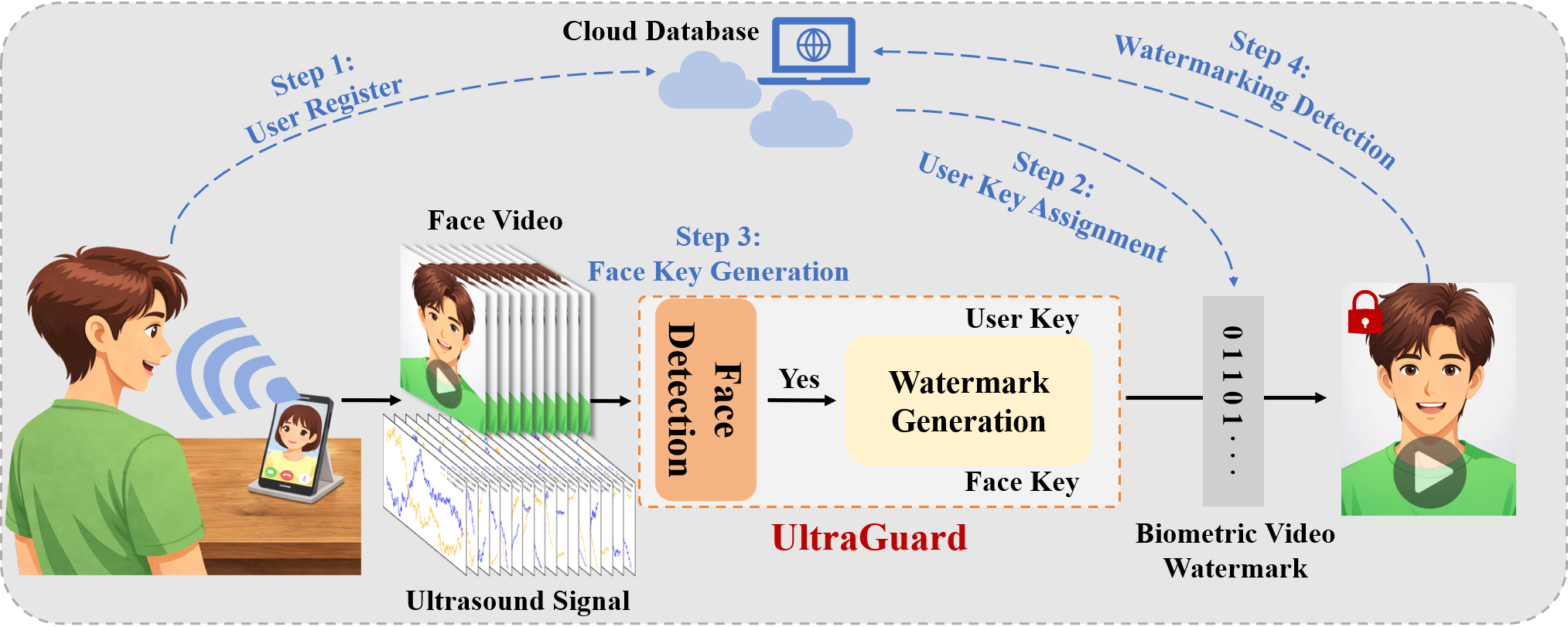}
  \caption{System overview of UltraGuard.}
  \label{fig: UltraGuard processing}
  \vspace{-0.2in}
\end{figure}

To make matters worse, existing methods address these threats in isolation. Video forgery detection approaches \cite{qian2020thinking, tariq2018detecting, corvi2023detection} focus solely on identifying post-hoc forgeries but provide no proactive protection at the source. Face authentication methods \cite{zhao2003face, schroff2015facenet, chen2019echoface} verify user identity at capture time but fail to protect video content once it enters the propagation channel. Neither approach provides an end-to-end security guarantee covering both generation and propagation. This fragmentation creates a critical gap where a video that passes authentication at capture can still be manipulated during transmission without detection. The growing deployment of video communication in high-stakes domains---such as remote judicial proceedings, telemedicine consultations, and executive business negotiations---further amplifies the urgency for a comprehensive solution that provides verifiable provenance from capture through the entire video content lifecycle.

In this paper, we present UltraGuard, a secure online dynamic watermarking system that leverages biometric information and facial content to defend face videos against deepfake attacks during both generation and propagation, as shown in Fig. \ref{fig: UltraGuard processing}.
The key insight is that source-side online dynamic watermarking is more secure by design than post-generation watermarking, because it grants valid watermark protection only to videos from authenticated live users and exposes unauthenticated or incorrectly watermarked videos as forgeries.
Our system offers two key advantages over traditional video watermarking systems.
First, our system is the first to leverage wireless reflections to capture face-related biometric information for secure online video watermarking, enabling live-user verification and preventing replay attacks that inject valid watermarks into forged videos at the source.
Second, our system is the first to introduce dynamic biometric video watermarking for secure video propagation, leveraging the uniqueness of facial changes within a time window to jointly encode user identity, facial motion dynamics, and biometric information, thereby significantly reducing the risk of malicious video tampering.

\begin{table*}[htbp]
  \centering
  \caption{Comparison between UltraGuard and existing video protection methods.}
  \scalebox{0.88}{
  \begin{tabular}{ccccccc}
    \toprule
    Method & \makecell{Source-side \\ Liveness} & \makecell{Protects \\ Propagation} & \makecell{Dynamic \\ Content Binding} & \makecell{Resistant to \\ Editing} & \makecell{Resistant to \\ Replay} & \makecell{Online \\ Deployment} \\
    \midrule
    \makecell{Video Forgery Detection\\ \cite{qian2020thinking, tariq2018detecting, corvi2023detection}} & \xmark & \xmark & \xmark & Limited & \xmark & \xmark\\
    \makecell{Traditional Video Watermarking\\ \cite{zhang2011spread, fernandez2024video, hu2025videomark}} & \xmark & \cmark & \xmark & Limited & \xmark & \xmark\\
    \makecell{Face Authentication\\ \cite{zhao2003face, schroff2015facenet, chen2019echoface}} & \cmark & \xmark & \xmark & \xmark & Partial & \cmark\\
    \textbf{UltraGuard} & \cmark & \cmark & \cmark & \cmark & Strong & \cmark \\
    \bottomrule
  \end{tabular}
  }
\label{tab: compare other methods}
\vspace{-0.2in}
\end{table*}

Despite its promise, several challenges must be addressed to realize a fully functional system.
%
(1) How to defend against source-side attacks while maintaining generalizability to newly registered users? We directly extract biometric features from ultrasonic waveforms and employ identity-aware facial video representations as auxiliary supervision for cross-modal alignment. Specifically, a large-scale pre-trained facial video encoder provides stable guidance to construct a discriminative ultrasonic feature space, where features from different users are well separated while remaining generalizable to unseen users, enabling reliable defense against source-side attacks.
(2) How to effectively defend against malicious attacks such as forgery and editing, thereby ensuring security during video propagation? By combining dynamic, time-window-adjustable encoding with spatial, real-time face detection, we construct a biometric video watermark based on spatio-temporal dynamic encoding. This dynamic encoding is derived from features obtained from our trained unimodal and fusion modal and exhibits strong correlation with facial information of different users. When a video containing this dynamic encoded watermark propagates over the internet, any malicious forgery or editing by an attacker creates a detectable discrepancy between the newly generated watermark and the original, correct one.
%
(3) How to deploy our system on resource-constrained smartphone devices while adapting to real-time video call scenarios also remains a key challenge. By analyzing each module of the watermarking system and integrating system optimizations such as dynamic face detection, key region localization, and a sliding-window cache pool, together with lightweight techniques like model quantization, we successfully achieve on-device deployment on smartphones while preserving the effective accuracy of the system, thereby embedding biometric video watermarks into face videos in real time.
The main contributions of this work are summarized as follows.

$\bullet$ To the best of our knowledge, UltraGuard is the first real-time online dynamic security watermarking system for facial videos. By jointly encoding user biometric information and facial content into time-varying watermarks, it enables source-side protection against forged video generation and propagation-stage defense against tampering and editing.

$\bullet$ To bridge the gap between fundamental insights and practical application, we address three key challenges: directly extracting features from users' biometric ultrasound signals and using video features for cross-modal alignment to build a discriminative, generalizable ultrasound feature space; constructing a face-correlated biometric video watermark via spatio-temporal dynamic encoding and fused video-ultrasound features for propagation security; and adapting the system to edge-side deployment using a range of optimizations and lightweight technique.

$\bullet$ We deploy our system on four smartphones and build a dataset with 50 volunteers, including 25 hours of legitimate samples and over 135 hours of attack samples across different source and propagation attack scenarios, achieving average 99.35\% and 100\% accuracy, respectively, which validates the effectiveness of our system.

\vspace{-0.1in}
\section{Related Work}
\subsection{Video Watermarking}

Video watermarking technology embeds imperceptible information into video content for purposes such as copyright protection, source tracing, and content authentication. Existing research primarily focuses on balancing robustness and imperceptibility, encompassing methods like spatial domain watermarking \cite{zhang2011spread}, transform domain watermarking \cite{langelaar2001optimal}, and temporal domain watermarking \cite{setyawan2001low}. Each of these approaches has inherent limitations on the robustness of the watermark or the quality of the video. 
In recent years, deep learning-based methods \cite{fernandez2024video, hu2025videomark, hu2025videoshield, xu2025invismark} have made significant progress. Researchers have designed end-to-end networks that jointly optimize the watermark embedding and extraction processes, demonstrating superior performance against complex distortions (e.g., transcoding, rescaling). 
This offers a new pathway towards achieving adaptive and highly robust video watermarking \cite{Luo2023DVMark, ye2023itov, cedillo2026deep}. However, current video watermarking techniques still mainly focus on static watermarking methods in offline scenarios. They cannot verify the authenticity of the video source nor guarantee sensitivity to forgery and tampering during propagation. Consequently, they are not widely applicable for real-time, online user scenarios, which represents a fundamental distinction from our proposed method.

\vspace{-0.1in}
\subsection{Video Forgery Detection}

With the rapid advances of deepfake, video forgery detection becomes a central problem in multimedia security. Existing methods mainly exploit two types of cues. One line of work examines biological signals and physical consistency, such as abnormal facial expressions, unnatural eye blinking, and inconsistencies in lighting, perspective, or physical reflections \cite{li2020identification, yang2019exposing}. These methods leverage real-world physical priors and work well against early forgery techniques, but their effectiveness declines as generative models produce increasingly realistic videos. To detect more convincing forgeries, recent studies adopt data-driven methods that train convolutional neural networks or Vision Transformers on large-scale fake-real datasets to capture subtle artifacts in pixel, frequency, or noise patterns \cite{qian2020thinking, tariq2018detecting}. More recently, several methods further target model-specific ``fingerprints'', such as traces left by diffusion models \cite{corvi2023detection}. However, existing detectors still suffer from limited generalization, and their performance often drops sharply when confronting unseen forgery methods. As AIGC continues to improve, purely detection-based defenses become increasingly fragile. This trend makes digital watermarking a more practical and promising direction for securing videos.

\vspace{-0.1in}
\subsection{Human Authentication Based on Smartphone}

Smartphone-based authentication includes 3D/infrared \cite{zhao2003face,rattani2018survey} and iris recognition \cite{daugman2009iris}, but they suffer from high cost, lighting constraints, or strong user intrusion.
Ultrasonic schemes using built-in speakers/microphones \cite{chen2019echoface,michalevsky2014gyrophone} are low-cost, light-immune, and unobtrusive for real-time communication.
However, pure authentication only discriminates users without carrying security information or maintaining continuity.
We advance beyond authentication by fusing ultrasonic biometrics with video watermarking to generate a dynamic, face-bound biometric watermark, enabling end-to-end security throughout video generation and propagation.

\vspace{-0.1in}
\subsection{UltraGuard Difference}

The fundamental logic of our watermarking system is to prevent source forgery at the origin through biometric authentication, while simultaneously achieving effective protection during video propagation by employing the proposed dynamic watermark correlated with video content, in conjunction with our detection system. Only facial videos carrying the correct user-specific and content-dependent watermark information are considered valid. Therefore, any forged video that lacks a watermark or contains an incorrectly encoded watermark can be identified as non-target content. This mechanism effectively protects every facial video generated by the user.

Compared to prior video protection schemes, our approach exhibits distinct differences, as shown in Table \ref{tab: compare other methods}.
Compared to video forgery detection schemes \cite{qian2020thinking, tariq2018detecting, corvi2023detection}, our method demonstrates overwhelming advantages, encompassing the entire process from the source to the propagation stage. Detection-based approaches only provide passive, one-time identification of forgeries and suffer from poor generalization.

Compared to face authentication schemes \cite{zhao2003face, schroff2015facenet, chen2019echoface}, UltraGuard overcomes their inherent limitation of only verifying identity at acquisition time. It enables authenticity traceability of videos during subsequent propagation by embedding dynamic content-related watermarks. Unlike authentication methods that only verify identity, carry no additional information, and lack long-term security, UltraGuard provides dynamic watermark protection throughout the entire propagation cycle by modifying the original video while preserving its visual content. 

Compared to traditional video watermarking \cite{zhang2011spread, fernandez2024video, hu2025videomark} schemes, UltraGuard supports source authenticity verification by binding source-side liveness information to propagated videos. When users transmit videos through remote devices, traditional watermarks cannot verify whether the content originated from a genuine live user. Attackers can exploit replay attacks to compromise the authenticity of traditional watermarks. By actively adding watermarks at the source and using dynamic encoding, UltraGuard overcomes these limitations.

In summary, unlike all previous methods, we propose a biometric video watermarking system that integrates with the biometric characteristics of target users, enabling protection from source to subsequent dissemination for all facial videos emitted by the target human.

\section{Preliminary}\label{sec: ultrasound preliminary}

In this section, we first present the background knowledge of ultrasound and video acquisition, and then introduce the temporal correspondence between ultrasound and video signals.

\vspace{-0.1in}
\subsection{Ultrasound and Video Acquisition}

We use the built-in top earpiece speaker to transmit ultrasound and the adjacent microphone to record audio at 48 kHz, while the front camera simultaneously captures facial video at 30 FPS. This co-located design makes the acoustic sensing field largely overlap with the visual capture region, enabling consistent observation of articulatory kinematics across acoustic and visual modalities~\cite{zhou2021robust}. We transmit a Continuous Wave (CW) signal, where $S(t)$ is a normalized superposition of $N=16$ orthogonal sub-carriers, i.e., $S(t) = \frac{1}{N} \sum_{k=0}^{N-1} \cos(2\pi f_k t)$, with $f_k$ ranging from 17.15 kHz to 22.4 kHz at 350 Hz intervals. Non-rigid articulatory movements perturb the acoustic paths and encode kinematic information into the received signal $R(t)$. To recover baseband motion information, we perform digital coherent demodulation on $R(t)$ using the locally generated in-phase carrier $\cos(2\pi f_k t)$ and quadrature carrier $-\sin(2\pi f_k t)$ for each $f_k$. We then apply a 50 Hz low-pass filter to extract baseband articulatory components and downsample the signals by $D=16$ to reduce downstream computation while preserving fine-grained motion details~\cite{lu2018lippass}. This process yields complex baseband signals at an effective sampling rate of 3 kHz, defined as $B_k(t) = I_k(t) + jQ_k(t)$, where $I_k(t)$ and $Q_k(t)$ capture the instantaneous motion-induced in-phase and quadrature components~\cite{wang2016device}.

\begin{figure}[t]
  \centering
  \includegraphics[width=0.7\linewidth]{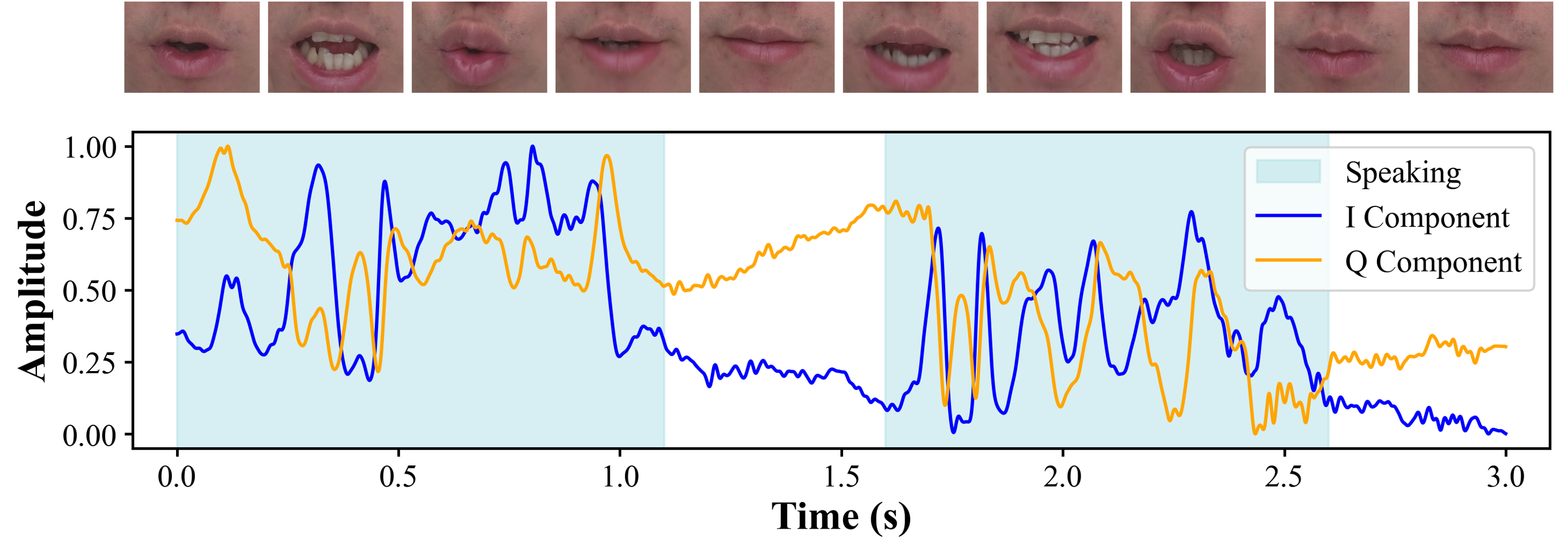}
  \caption{Ultrasound-video Synchronization.}
  \label{fig: synchronization}
  \vspace{-0.2in}
\end{figure}

\vspace{-0.1in}
\subsection{Ultrasound-video Consistency}

To verify the intrinsic coupling between ultrasonic signals and visual actions, we analyze the correlation between concurrently captured video and acoustic features. As shown in Fig. \ref{fig: synchronization}, speech video frames are aligned with the corresponding In-phase $I_k(t)$ and Quadrature $Q_k(t)$ components of the 17.15 kHz sub-carrier along a shared timeline, allowing us to examine the synchronous effect of mouth movements on both modalities.
The results show strong consistency between the fluctuations of $I_k(t)$ and $Q_k(t)$ and the mouth movements in the video. During silent intervals, the acoustic features remain relatively stable because the mouth region undergoes little physical deformation. Once the user starts speaking, both $I_k(t)$ and $Q_k(t)$ exhibit clear fluctuations. This result demonstrates that ultrasound reliably captures fine-grained articulatory kinematics and preserves temporal synchronization with the video, providing a solid basis for subsequent dynamic watermarking.

\vspace{-0.1in}
\section{Background and Threat Model}
\subsection{Design Goals}

UltraGuard aims to preserve facial video authenticity from generation to propagation. Unlike prior systems that address either source authentication or content protection in isolation, UltraGuard seeks to unify both properties into a single, practical framework. Based on this objective, we define three design goals.
\textbf{Goal 1: Source Authenticity.}
The system should verify that a facial video originates from a legitimate live user and resist replay, face-swapping, and video forgery attacks at the source. This requires the system to establish live-user evidence that cannot be fabricated from pre-recorded content alone. Our approach leverages the unique physical coupling between the user's face and the ultrasonic signal reflected from it, creating a biometric signature that is inherently tied to the live presence of the enrolled user.
\textbf{Goal 2: Propagation Integrity.}
The system should detect unauthorized modifications to propagated facial videos, including fine-grained editing, cropping, and splicing operations on face content. Unlike static watermarks that remain constant regardless of content, the integrity protection must be content-dependent, such that any modification to the video's facial content produces a detectable change in the watermark verification result. This ensures that even surgically precise edits to small facial regions are detectable.
\textbf{Goal 3: Practical Deployment.}
The system should provide content-aware authenticity protection with low latency and robustness under diverse real-world conditions, including varying distances, angles, ambient noise levels, and device types. Furthermore, the system must be deployable without requiring specialized hardware beyond standard smartphone components, and should support efficient adaptation to edge devices through lightweight optimization techniques.

\subsection{Threat Model}\label{sec: threat model}
\subsubsection{System Assumptions}

UltraGuard aims to preserve facial video authenticity throughout capture, transmission, and subsequent propagation. Its security relies on establishing a trusted association between the user's facial appearance captured by the front camera and the corresponding biometric-related signals captured through the smartphone's audio subsystem. This association is encoded into a biometric watermark that accompanies the video as it propagates across the Internet.

We assume that the user's smartphone, including the microphone, earpiece speaker, front camera, operating system, and UltraGuard's sensing and execution components, operates correctly and is not compromised. Sensor measurements are acquired through the device's normal sensing interfaces. In particular, the adversary cannot directly tamper with sensor hardware or drivers, bypass the sensing pipeline to inject fabricated digital measurements, or otherwise manipulate the trusted sensing process. These assumptions define the trusted hardware and sensing boundary of UltraGuard. Attacks that compromise this boundary are outside the scope of our watermarking system.

The ultrasonic evidence used by UltraGuard is generated by the smartphone's own speaker and captured by its microphone through the normal acoustic propagation path. The acquisition and binding of biometric-related ultrasonic evidence are performed within the trusted sensing pipeline. We assume that the raw ultrasonic measurements and the internal correspondence between the target user's biometric characteristics and the captured ultrasonic responses are protected from direct adversarial access. In particular, the adversary cannot obtain target-specific paired biometric-ultrasound observations sufficient to reconstruct the correspondence exploited by UltraGuard. This assumption concerns the confidentiality of the trusted acquisition and binding process, rather than the physical impossibility of external ultrasonic interference.

The cloud verification service is trusted and follows the prescribed protocol. Secure communication between trusted components can be established through standard transport-layer security mechanisms. We also assume trusted user enrollment, following the standard assumption adopted by biometric authentication systems. As with Face ID and fingerprint authentication, the security of UltraGuard fundamentally depends on the correctness of the enrolled identity. An adversary who impersonates the victim during enrollment compromises the root of trust rather than exploiting a vulnerability of UltraGuard itself. Moreover, attacks against UltraGuard are meaningful only after a user has enrolled and enabled watermarking. In practice, enrollment can leverage trusted mechanisms already available on modern smartphones, such as Face ID, fingerprint authentication, or trusted execution environments. Enrollment attacks and full compromise of trusted hardware or software are therefore outside our threat model.

\subsubsection{Attack Model}


We consider a strong adversary who knows the general working principle the UltraGuard architecture. The adversary may obtain previously published facial videos, corresponding watermarked videos, and access to the static detector implementation. Notably, static detectors differ from our cloud-based dynamic detectors in that they can only detect the statically assigned User Key described later, consistent with watermark detection schemes in most prior studies. In contrast, verification of the dynamic watermark (Face Key) is performed exclusively within our trusted cloud service, and the associated secret verification material is not disclosed to the adversary.

The adversary can perform replay attacks, face-swapping, video editing, watermark reuse, and other digital manipulations to bypass the verification process. We further consider adaptive adversaries who are aware of the multimodal consistency constraints exploited by UltraGuard. Such adversaries may attempt to construct synchronized forged videos and ultrasonic signals to maximize their acceptance probability.

However, these attacks are constrained by the trusted sensing and binding assumptions described above. The adversary cannot directly inject fabricated measurements into the smartphone's sensor data stream or access the internal biometric-to-ultrasound correspondence used by UltraGuard. Attacks that require eavesdropping on the target user's ultrasonic responses or collecting sufficient target-specific paired biometric-ultrasound observations to train a model that reproduces this correspondence are therefore outside our threat model. Similarly, the adversary is assumed not to have access to synchronized video-ultrasound pairs collected during legitimate watermark generation when such access would expose the protected correspondence. These restrictions do not imply that external ultrasonic signals are physically incapable of reaching the microphone; rather, they delimit the adversary's access to the trusted measurements and target-specific information used by the verification process.

Accordingly, UltraGuard focuses on protecting the association between legitimately captured biometric signals and facial video, and on detecting forged or manipulated content as it propagates across the Internet. Direct physical attacks on trusted sensors, malicious modification of sensor drivers or firmware, privileged compromise of the operating system, and other attacks that violate the trusted sensing boundary are outside the scope of this work.
Future improvements may further reinforce the trusted sensing pipeline through hardware-level encryption and data-flow integrity verification, reducing the risk of unauthorized access to internal sensor data and malicious data injection.



\section{System Overview}

The overall design logic of the system is that only facial videos carrying the correct dynamic watermark information are considered authentic representations of the user. Videos lacking a watermark or containing incorrect watermark information are deemed fake. The system effectively defends against forgery attempts at the source, while our dynamic watermark encoding provides robust protection against forgery or editing of acquired user facial videos during transmission.

The UltraGuard can be effectively deployed on the mobile devices of different users in the future, providing online, real-time protection for their facial videos and preventing forgery both at the source and during propagation. Its overall workflow, illustrated in Fig. \ref{fig: UltraGuard processing}, consists of the following main steps:
1. \textbf{User Registration.} Before using the system, the user registers by providing a sample of facial video features and the corresponding ultrasonic features. 
Specifically, the registered samples are converted into compressed feature vectors by the model output, stripped of the user’s biometric information, and then uploaded to a trusted cloud for subsequent
watermarking and detection.
%
2. \textbf{User Key Assignment.} Upon successful registration, the system automatically assigns a User Key code. This key will later be combined with the Face Key to form the dynamic watermark embedded into the video.
3. \textbf{Face Key Generation.} During an actual video call, the system automatically and dynamically generates a corresponding Face Key based on the live facial video content. This key is used for dynamic encoding, enabling effective, real-time protection.
4. \textbf{Watermark Detection.} When the receiving end needs to verify whether the current video has been forged or edited by an attacker, the video is transmitted to our system. The system extracts the encoding information from the video and compares it with the encoding information derived from the video itself. A mismatch indicates that the original video has been modified.

After registering with UltraGuard, whenever a user engages in any video call on their mobile phone, they simply need to activate the UltraGuard system. This automatically adds a unique, dynamic watermark to the outgoing video signal while maintaining imperceptible transmission that does not affect the original video quality.
Furthermore, the system incorporates face detection functionality. When no face is detected, the ultrasonic sensor is not activated, and no watermark is added to the video. 
On the detection end, for simple scenarios, the detector can verify the video using the User Key, similar to current mainstream static watermarking techniques.
For facial videos that an attacker has meticulously modified or edited, the video is uploaded to our system. The system generates the corresponding watermark information based on the current video content, compares it with the Face Key watermark information embedded within the video, and returns the comparison result, thereby determining whether the video has been tampered with or forged.
Benefiting from the ubiquity of the microphone and earpiece speaker required for ultrasound in all smartphones across all price segments, UltraGuard possesses significant generalizability and broad applicability without requiring hardware modifications. Moreover, the fast single-sample enrollment process makes it practical for large-scale user adoption in real-world deployment scenarios.

\section{System Design}

We need to integrate video and ultrasound features, but the gap between the two modalities is large. Therefore, we adopt a late-fusion strategy to combine them.
The advantage of late fusion is twofold. First, it facilitates the effective use of a pre-trained video encoder. Second, and more importantly, it mitigates modality dominance. Since the video modality contains substantially richer information than the ultrasonic modality, direct fusion can cause video features to overwhelm ultrasonic features and hinder ultrasonic training.
Specifically, the video and ultrasonic modalities are first mapped into a unified feature space by their respective encoders. This design places them at a comparable representation level and reduces mutual suppression. Fusion and subsequent processing are then performed on these aligned features.

We introduce the system in the following four parts.
First, we present the generalization-oriented design of the ultrasonic encoder to ensure source security.
Second, we explain the generation of the spatio-temporal dynamic encoding for video watermark features and the system decision logic.
Next, we present the design of the video encoder, focusing on the rationale behind the feature representation and the motivation for incorporating a pre-trained model.
Finally, we describe watermark embedding into the original video and subsequent detection, along with edge deployment strategies.

\begin{figure}[t]
  \centering
  \includegraphics[width=0.8\linewidth]{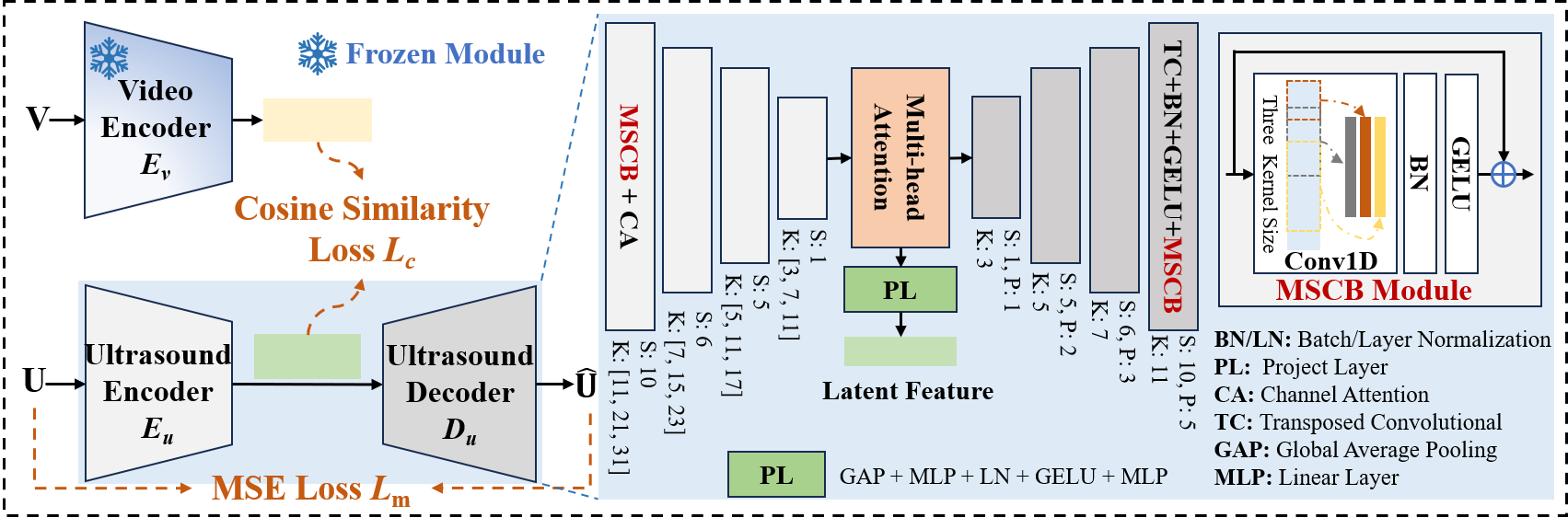}
  \caption{The architecture of the ultrasound encoder and decoder.}
  \label{fig: model overview_stage1}
  \vspace{-0.2in}
\end{figure}

\subsection{The Design of Ultrasound Encoder}\label{sec: ultrasound encoder}
\subsubsection{The Design of Generalization}

The primary objective of this ultrasound encoder $E_u$ is to distinguish between the input features of legitimate users and those of malicious attacks at the source. Therefore, it must ensure that the single enrollment sample of a user is effectively clustered with other samples from the same user. Consequently, the main goal of $E_u$ is to project features from different users into a stable, high-dimensional space where clear separation exists between distinct users.
As shown in the left part of Fig. \ref{fig: model overview_stage1}, we use $E_u$ to encode the ultrasonic signal (U), and use a mean squared error (MSE) loss $L_m$ along with a weak ultrasonic signal decoder $D_u$ to reconstruct the corresponding ultrasonic signal, ensuring the validity of the learned intermediate features.
This design constitutes an adversarial training scheme in which the weak decoder encourages the model to preserve intra-class variability while learning discriminative inter-user boundaries, thereby forming well-separated user clusters without causing within-class feature collapse.

To prevent feature loss from the ultrasonic signal, we directly process its raw waveform.
But this approach of training an encoder via an encoder-decoder architecture imposes high demands on data volume. 
If trained solely on the ultrasonic data collected from our limited number of users, the resulting model's ability to generalize to ultrasonic signals from unseen users would be severely limited.

Therefore, we employ a mature video encoder $E_v$ (detailed in Section \ref{sec: video encoder}), which was trained on a large-scale facial dataset, to encode the video (V) modality for simultaneous training with $E_u$.
Since $E_v$ is built upon a large-scale pre-trained model, it can form a large capacity, good generalization, and stable feature space.
By leveraging the robust feature space provided by a mature video encoder and aligning the ultrasonic modality within it through feature embedding, we can significantly expand the expressive capacity of ultrasonic features and enhance generalization capability \cite{radford2021learning, kimura2025infomae}. This approach yields substantially better results compared to encoding the ultrasonic signal in isolation.
Consequently, the feature space learned by our ultrasonic signal encoder $E_u$ is not limited to the finite ultrasonic samples we collected. 
It ensures that when a new user registers, the region assigned to their features in the feature space is sufficiently distinct from those of other users, which enhances our system's generalization for distinguishing source malicious attacks and normal users.
We train $E_u$ using a cosine similarity loss function $L_c$ to align its output with the output features of $E_v$. The formula is as follows:
\begin{equation}
\mathcal{L}_{\text{c}} = 1 - \frac{\mathbf{a}^\top \mathbf{b}}{\|\mathbf{a}\|_2 \cdot \|\mathbf{b}\|_2}
\end{equation}
Where $\mathbf{a}$ and $\mathbf{b}$ represent the predict and label vector respectively, and $\|\cdot\|_2$ is the 2-norm. 
The overall loss function of this stage is as follows, regulated by the coefficients $\lambda_c$ and $\lambda_m$:
\begin{equation}
    \mathcal{L}_{\text{s1}} = \lambda_c \mathcal{L}_{\text{c}} + \lambda_m \mathcal{L}_{\text{m}}
\end{equation}
By aligning the video and ultrasonic features, we enable the ultrasonic encoder to acquire cross-semantic capabilities.

\subsubsection{The Structure of Ultrasound Encoder}

As shown in the right part of Fig. \ref{fig: model overview_stage1}, we construct an encoder-decoder model that integrates multi-scale feature extraction and attention mechanism to process sequential ultrasonic waveform signals. The model adopts a symmetrical encoder-decoder architecture, with its core design focused on multi-scale capture of temporal features and reinforcement of key information. 
The encoder takes multi-channel, variable-length ultrasonic waveform signals as input. It employs four cascaded multi-scale convolutional blocks (MSCB) for progressive down-sampling. Each block integrates three parallel convolutions with different kernel sizes to capture both fine-grained and global temporal features. 
A channel attention (CA) mechanism \cite{wang2020eca} is also introduced to weight key feature channels while suppressing redundant information,
and finally projects the features to a feature map.
This feature map is transposed in dimension and fed into an attention block with positional encoding, which injects temporal positional information and achieves feature fusion through multi-head attention and feed-forward networks. The output is a latent feature with dimension 512 (a linear layer is added to the output of $E_v$ to transform its dimensionality, aligning it with the target dimension).
To separate feature learning from waveform reconstruction, the latent features undergo global average pooling (GAP) and are then fed into a projection layer (PL) to generate fixed-dimensional features for downstream tasks.
The decoder uses four cascaded transposed convolutional (TC) blocks for progressive up-sampling (symmetrical to the encoder's down-sampling), ultimately outputting a reconstructed waveform with dimensions matching the input.
Furthermore, we incorporate a model embedding dimension padding mechanism to ensure temporal signal dimensions remain consistent for any value of T, providing an effective guarantee for dynamic encoding over arbitrary time intervals.

\begin{figure}[t]
  \centering
  \includegraphics[width=0.5\linewidth]{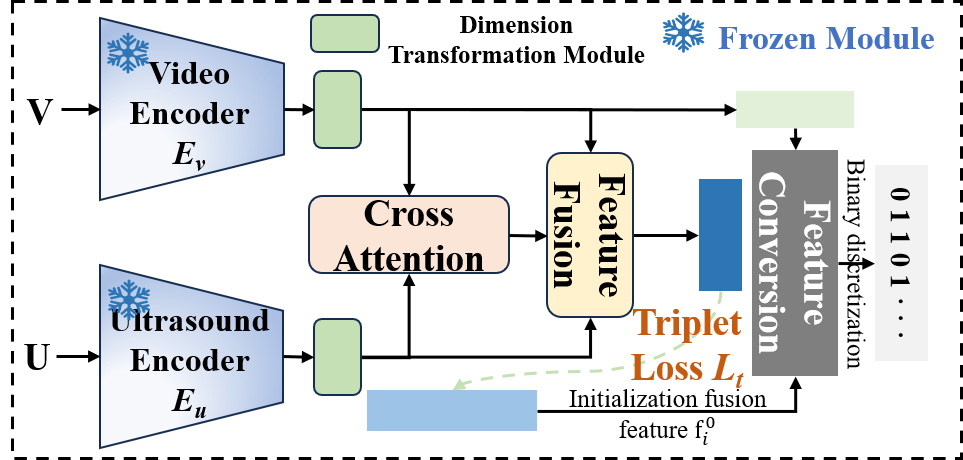}
  \caption{The architecture of fusion feature module.}
  \label{fig: model overview_stage2}
  \vspace{-0.2in}
\end{figure}

\vspace{-0.1in}
\subsection{User Verification and Watermark Generation}\label{sec: watermarking generation}
\subsubsection{Fusion Feature Generation}

The primary objective of this stage is to train the fusion of feature vectors from $E_v$ and $E_u$, which are subsequently used for watermark encoding and multi-level user verification, where $E_v$ is the mature facial video encoder in Section \ref{sec: video encoder}.
As shown in Fig. \ref{fig: model overview_stage2}, the output features from the two encoders are passed through a dimension transformation module (DTM) to convert them to a target encoding dimension. These two features are then fused via a cross-modal cross-attention module. Finally, the combined representation, along with the two modal features, is fed into a feature fusion module for transformation, yielding the fused feature vector $f_i^0$.

To enable effective dimension transformation for newly registered user vectors and to allocate the fused feature vectors to appropriate regions in the fused feature space, we employ the Triplet Loss $L_t$ for metric learning:
\begin{equation}
    \mathcal{L}_{\text{t}} = \max \left( \|\mathbf{x}_a - \mathbf{x}_p\|_2^2 - \|\mathbf{x}_a - \mathbf{x}_n\|_2^2 + \alpha, 0 \right)
\end{equation}
Where $\mathbf{x}_a$ is the feature of the anchor point, $\mathbf{x}_p$ is the positive sample feature same as the user of the anchor point, $\mathbf{x}_n$ is the negative sample feature different from the user of anchor point, and $\alpha$ is the margin.
Benefiting from the two modal encoders trained in the first stage, the modules trained in this stage can easily cluster the effective sample pairs corresponding to the two modalities of the same user, while effectively separating clusters of different users.

The DTM consists of two fully connected layers followed by GELU activation, with the number of neurons in the final layer matching the subsequent encoding dimension (256). The cross-attention module primarily fuses features from the two modalities along the temporal dimension, based on a multi-head attention mechanism with 8 heads. The fusion module is composed of three sequential blocks, each containing a linear layer, Batch Normalization, and a GELU activation layer.

\subsubsection{User Verification}

The logic for determining whether a user is the target involves comparing the user's video feature $\mathbf{v}_i^j$, ultrasonic feature $\mathbf{u}_i^j$, and fusion feature $\mathbf{f}_i^j$ against the initial enrollment vectors stored in our system, where $i$, $j$ represent the index of user and the corresponding sample feature, respectively. The decision process first requires the similarity of the $\mathbf{v}_i^j$ feature to exceed a threshold $\beta_v$. Only then are the ultrasonic and fusion features evaluated. The user is identified as the current target only if the similarities of both $\mathbf{u}_i^j$ and $\mathbf{f}_i^j$ exceed their respective thresholds $\beta_u$ and $\beta_f$. We set $\beta_v$, $\beta_u$, and $\beta_f$ to 0.5.

\begin{algorithm}[t]
\caption{Target User Verification.} \label{alg:user_verification}
\begin{algorithmic}[1]
\Require Current feature vectors: video feature \(\mathbf{v}_i^j\), ultrasound feature \(\mathbf{u}_i^j\), fused feature \(\mathbf{f}_i^j\)
\Require Enrolled reference vectors: \(\mathbf{v}_i^0\), \(\mathbf{u}_i^0\), \(\mathbf{f}_i^0\)
\Require Similarity thresholds: \(\beta_v\), \(\beta_u\), \(\beta_f\)
\Ensure Verification result: True (Accept) or False (Reject)
\State \(S_v \leftarrow \text{Similarity}(\mathbf{v}_i^j, \mathbf{v}_i^0)\)
\If{\(S_v < \beta_v\)}
    \State \Return{False}
\EndIf
\State \(S_u \leftarrow \text{Similarity}(\mathbf{u}_i^j, \mathbf{u}_i^0)\), \(S_f \leftarrow \text{Similarity}(\mathbf{f}_i^j, \mathbf{f}_i^0)\)
\If{\(S_u \geq \beta_u\) and \(S_f \geq \beta_f\)}
    \State \Return{True}
\Else
    \State \Return{False}
\EndIf
\end{algorithmic}
\end{algorithm}

\subsubsection{Dynamic Biometric Video Watermarking}

To ensure security for video propagation, we propose a spatio-temporal dynamic encoding-based biometric video watermark.
The watermark is strongly correlated with facial content, reflected in its dependence on the facial features within the video over a time window T. 
When facial content is edited or modified, changes in the watermark accordingly reflect the tampering. The temporal aspect is manifested in the adjustable window T. By balancing the time range and the computational capacity of the specific device, an appropriate value for T can be selected. 
The spatial aspect is reflected in the system's behavior: if a face is not detected within the camera's field of view, the UltraGuard system halts, ceasing both ultrasonic signal emission and video watermark generation. Besides, combined with subsequent edge-side optimization (Section \ref{sec: edge deployment}), the system focuses the watermark addition on the region of interest (ROI) of the face, thereby reducing the model's computational load and enabling effective edge-side deployment.

In the specific system workflow, after user registration, our system assigns the user a User Key, which consists of $U$ bits. This component is a fixed code, similar to traditional video watermarks, and is effective for simple or real-time security scenarios. 
For meticulously crafted attacks, we propose a more refined and content-dependent dynamic Face Key. We transform the facial video features within the time window T to obtain the corresponding video vector feature $\mathbf{v}_i^j$, which fuses face frames within that period. This vector encapsulates the semantic facial content during that interval, capable of reflecting fine-grained facial alterations.
Even for the same face, differences in expressions or other facial characteristics result in distinct vectors within the cluster, forming the basis for constructing our dynamic Face Key encoding of $F$ bits.

\begin{algorithm}[t]
\caption{Dynamic Encoding.}\label{alg:feature_transform}
\begin{algorithmic}[1]
    \Statex \noindent\textbf{Input:} Original video $\mathbf{V}_i^j$, ultrasound signal $\mathbf{U}_i^j$ of $j$-th sample of user $i$
    \Statex \noindent\textbf{Output:} Watermark code $\mathbf{B}_i^j$ of $j$-th sample of user $i$
    
    \State Registered initial video feature $\mathbf{v}_i^0$, ultrasound feature $\mathbf{u}_i^0$, and fusion feature $\mathbf{f}_i^0$ for user $i$, assign the User Key $UK_i$ to user $i$
    \State Extract video features $ \mathbf{v}_i^j \gets E_v(\mathbf{V}_i^j)$, ultrasound features $ \mathbf{u}_i^j \gets E_u(\mathbf{U}_i^j)$
    \State Feature dimension transformation: $\mathbf{v}_i^j \gets \text{DTM}(\mathbf{v}_i^j)$, $\mathbf{u}_i^j \gets \text{DTM}(\mathbf{u}_i^j)$
    \State Feature fusion: $\mathbf{f}_i^j \gets \text{FeatureFusion}(\mathbf{v}_i^j, \mathbf{u}_i^j)$
    \State Dynamic generate Face Key: $FK_i^j \gets  \text{sign}(\tanh(\text{norm}(\mathbf{v}_i^j 
 - \mathbf{v}_i^0)*\text{norm}(\mathbf{f}_i^0)))$
    \State Concatenate User Key $UK_i$ and Face Key $FK_i^j$: $B_i^j \gets Concat([UK_i,\ FK_i^j])$ 
\end{algorithmic}
\end{algorithm}

The specific dynamic encoding algorithm is outlined in Algorithm \ref{alg:feature_transform}. We subtract the registered video feature vector $\mathbf{v}_i^0$ from the dimension-transformed video feature vector $\mathbf{v}_i^j$, normalize the result, and then multiply it by the initially registered and normalized fusion vector $\mathbf{f}_i^0$. Finally, this vector is passed through tanh and sign functions to convert it into a dynamic binary watermark code (Face Key $FK_i^j$), which is concatenated with the static User Key $UK_i$ to form the final watermark code $B_i^j$. 
Because our encoding information is directly derived from high-dimensional feature space vectors, undergoes binarization via lossy compression, and is transformed by the fused feature vector obtained during user registration, it hardly be reverse-engineered.

The reason we do not directly use the user's real-time ultrasonic vector in the encoding is that at the detection end, the corresponding ultrasonic signal is unavailable, so the ultrasonic encoding vector cannot serve directly as the watermark content. Instead, we transform the encoding vector using the fused vector obtained during user registration. We set the Face Key $F$ to 224 bits and the User Key $U$ to 32 bits, totaling 256 bits.

\subsection{The Design of Video Encoder}\label{sec: video encoder}

Abstracting the most fundamental requirements from Section \ref{sec: ultrasound encoder} and Section \ref{sec: watermarking generation}, the video encoder must be capable of distinguishing between different users, ensuring the independence of their features to defend against source attacks, while simultaneously distinguishing between different facial information from the same user, essential for dynamic watermarking.

\begin{figure}[t]
  \centering
  \includegraphics[width=0.6\linewidth]{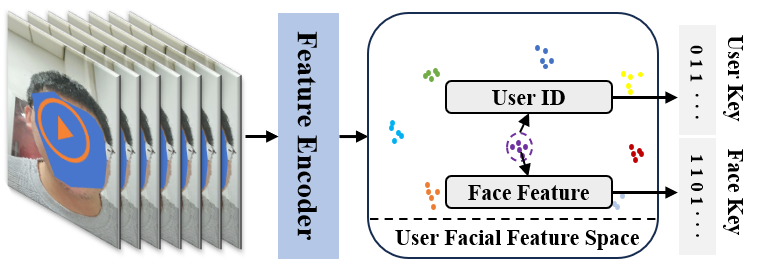}
  \caption{The video facial feature cluster for different registrant users.}
  \label{fig: cluster show}
   \vspace{-0.2in}
\end{figure}

Therefore, for the video encoder $E_v$, an effective feature space must be constructed where features from different users form clearly distinct clusters, while features from different facial videos of the same user remain differentiated within the boundaries of that user's cluster, as illustrated in Fig. \ref{fig: cluster show}.
Inspired by the face detection task \cite{zhang2016joint, najibi2017ssh, deng2020retinaface}, we observe that faces from different individuals naturally form clusters in the learned feature space. However, for the same individual, due to the semantic leakage of facial characteristics (e.g., expression, position), features from the same face do not converge to a single point. Instead, they are distributed within a sub-region of the cluster. While this is a drawback for face recognition, it aligns perfectly with the requirements of our task.

We utilize a pre-trained ArcFace \cite{deng2019arcface} model,  as our video encoder $E_v$ (shown in the left part of Fig. \ref{fig: model overview_stage1} and Fig. \ref{fig: model overview_stage2}), to construct the facial identity feature space. By introducing an additive angular margin during training, ArcFace projects facial images into a normalized hypersphere, where features from the same identity are tightly clustered while those from different identities are well separated. 
This angular-margin optimization enables the extracted facial embeddings to remain stable under variations in pose, illumination, and facial expression, making them well suited for identity verification tasks.
Besides, such a highly discriminative and semantically meaningful embedding space provides reliable identity supervision for cross-modal feature alignment with the ultrasonic modality. 
In our implementation, each video frame is processed by the pre-trained ArcFace network to obtain a 512-dimensional identity embedding, and the frame-level features are subsequently aggregated to form a robust video-level facial representation.

\subsection{Watermark Embedding and Detection}

After obtaining our dynamic video watermark, it must be seamlessly integrated into the user's facial video. The primary objectives are to ensure the watermark is imperceptibly embedded without degrading video quality, and to maintain the watermark's sensitivity to malicious attacks and robustness against benign modifications during the video's propagation.
%
%
We employ InvisMark \cite{xu2025invismark} to embed identity-related watermarks into the facial video. InvisMark is an end-to-end neural watermarking framework comprising a watermark encoder and a blind decoder, where the encoder imperceptibly injects a binary watermark into the input image and the decoder recovers the embedded message directly from the received image without referencing the original content.
To enhance robustness, the training process incorporates a differentiable distortion layer that simulates common image processing operations, including compression, resizing, filtering, and geometric transformations, enabling the learned watermark to survive practical transmission and editing pipelines. 
Compared with conventional watermarking approaches, InvisMark supports high-capacity 256-bit payload embedding while simultaneously achieving excellent visual fidelity and robust decoding performance under diverse distortions. 
In our system, the watermark is embedded into the facial video before transmission, allowing subsequent integrity verification while introducing negligible visual artifacts.


The encoded information is strongly bound to the facial video content during that specific time window. When an attacker modifies any portion of the video, this holistic binding is disrupted, causing a discrepancy in the corresponding encoded information.
Through extensive adversarial training, InvisMark maintains robustness against compression losses incurred during transmission and propagation, enabling effective extraction of the watermark information.
Given a system-assigned fixed User Key, the detector can locally extract, compare, and verify the watermark, suitable for applications requiring low-to-moderate security.
For high-security scenarios, verification of the dynamic Face Key watermark is necessary: the detector uploads the video to the cloud, where our cloud-deployed UltraGuard (operating without biometric authentication) generates the corresponding watermark, compares it with the embedded Face Key, and returns the result.
Our detector supports dynamic configuration to adapt to diverse scenario-specific security requirements.
In addition, to enable deployment on resource-constrained smartphone end devices, we optimize and adapt the watermark embedding and detection process for on-device deployment, with specific details provided in follow section.

\begin{figure}[t]
  \centering
  \includegraphics[width=0.7\linewidth]{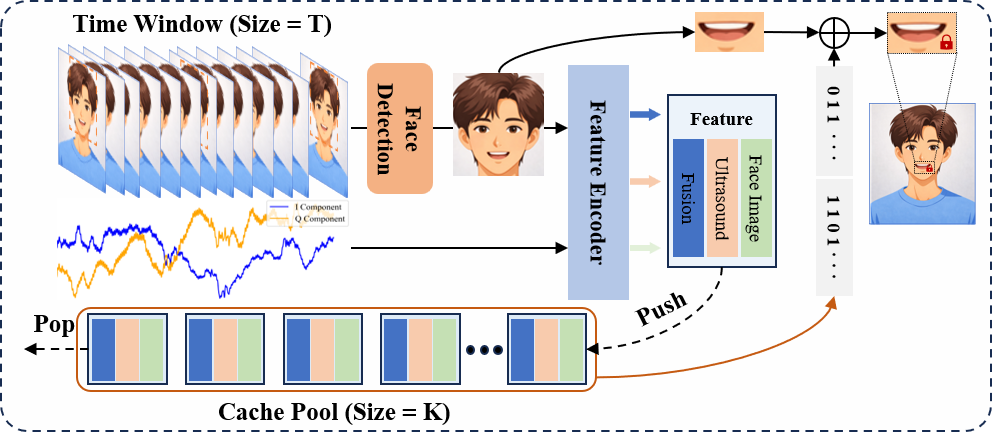}
  \caption{The video watermarking flowchart using edge deployment optimization.}
  \label{fig: Edge deployment}
   \vspace{-0.2in}
\end{figure}

\subsection{Edge Deployment}\label{sec: edge deployment}

To enable practical deployment of UltraGuard on resource-constrained smartphone edge devices, we incorporate several lightweight optimization strategies, as shown in Fig. \ref{fig: Edge deployment}.






\subsubsection{Face Detection and Key Region Localization}

To enable real-time deployment on mobile devices, we first perform lightweight face detection on the input video stream. The detector continuously locates the face in each frame and serves as a trigger for the subsequent watermark embedding process. Specifically, watermark embedding is executed only when a valid face is detected; otherwise, the current frame is skipped. This dynamic pipeline naturally supports temporal watermarking while avoiding unnecessary computation on background or non-face frames.
Beyond face localization, we further exploit the facial landmarks returned by the detector to identify the lip region, which is the primary manipulation target of current facial forgery methods (e.g., lip-sync generation and talking-face synthesis). Instead of embedding watermarks into the entire facial image, we only operate on this key semantic region, significantly reducing computational overhead while preserving watermark robustness.
For face detection, we adopt SCRFD-500M, a lightweight face detector proposed in SCRFD \cite{guo2022scrfd}, which achieves an excellent accuracy-efficiency trade-off for edge devices.
Besides the face bounding box, the detector outputs facial landmark coordinates, including the two mouth corners and nose landmarks. We utilize the left and right lip corner coordinates to determine the mouth center, and then expand the region vertically according to a predefined aspect ratio to generate a fixed 256×256 ROI. This normalized ROI is subsequently resized as the input of the watermarking network, ensuring consistent spatial alignment across different users, poses, and camera distances.
Compared with watermarking the entire face, the proposed key-region localization substantially reduces the embedding area while concentrating watermark information on the most security-critical region, thereby improving deployment efficiency and reducing visual distortion.
Given that the front-facing camera video window of a conventional smartphone is approximately 1280×720 pixels, under normal video-calling scenarios (where the camera can fully capture the face), this region is sufficiently large to cover the lip area of the human face.

\subsubsection{Key Region Watermark Embedding}

Given the extracted face region, we first feed it into the feature encoder to extract high-level facial representations.
Specifically, we adopt the lightweight feature encoder, which employs a MobileFaceNet-based \cite{chen2018mobilefacenets} backbone trained with the ArcFace loss to extract 512-dimensional facial embeddings.
Compared with the original ArcFace model based on deeper backbones, the encoder employs a lightweight backbone, achieving a significantly smaller model size and lower computational cost while maintaining discriminative face embeddings through ArcFace training.
The encoded feature is further transformed into a dynamic Face Key watermark, whose content varies across different users and video frames. Consequently, each watermark is tightly bound to both the facial identity and temporal context, making replay or forgery attacks considerably more difficult.
Instead of modifying the entire face image, the generated watermark is embedded only within the localized lip region. This design is motivated by the observation that most state-of-the-art facial manipulation methods modify only local facial regions, particularly the mouth area, while leaving the remaining facial content largely unchanged. Embedding the watermark into the same manipulated region enables the watermark to undergo the same geometric and appearance transformations as the forged content, thereby significantly improving its survivability after face editing or lip synchronization.
After watermark embedding, the watermarked lip ROI is seamlessly fused back into the original facial image according to its spatial coordinates, producing the final protected video frame. Since only a small local region is modified, the proposed strategy effectively minimizes visual artifacts while reducing computational complexity, making it suitable for real-time mobile deployment.

\subsubsection{Sliding Window Cache Pool}
To further reduce computational latency on resource-constrained mobile devices, we observe that watermarking every video frame is unnecessary for reliable authentication. Since our verification logic only requires detecting a sufficient number of valid watermarks, videos lacking enough authenticated watermarks can be directly regarded as forged or unauthenticated. Therefore, sparse watermark embedding can preserve security while substantially improving efficiency.
We introduce a sliding-window cache pool to selectively process representative frames. Instead of watermarking every frame, we uniformly sample $K$ key frames within each temporal window and perform feature extraction and watermark embedding only on these frames. The extracted watermark features are cached and reused within the current window, avoiding redundant computation on adjacent frames with highly similar facial content.
In our implementation, the input video is recorded at 30 FPS, and each sliding window spans 1 second. We empirically set $K=5$, meaning that only five uniformly distributed frames are selected for watermark embedding per window, while the remaining frames bypass the embedding module. Consequently, the watermarking frequency is reduced by approximately 83.3\% (30$\rightarrow$5 frames), substantially lowering inference latency and energy consumption without sacrificing authentication reliability.
For each selected frame, its face feature is extracted and combined with the features of other frames in the cache pool to obtain an average feature. This average feature is then converted into watermark information and embedded into the current frame, associating it with the features of other key frames from the preceding 1 second. The current feature is subsequently pushed into the cache pool for subsequent key frames, while the oldest feature is popped out to maintain a fixed pool size. This sparse embedding strategy provides an effective trade-off between security and deployment efficiency, making it well suited for real-time mobile inference.

\subsubsection{Module Quantization}

%
%
To further reduce the model size and inference latency on mobile devices, we apply post-training quantization to the system components that have not been optimized previously. For these modules, we adopt INT8 quantization with Quantization-Aware Training (QAT) \cite{jacob2018quantization}, which mitigates the accuracy degradation introduced by direct quantization.
QAT works by inserting fake quantization modules during training to simulate INT8 computation. This allows the model to compensate for quantization errors during optimization. After training, the model is exported as a fully INT8 network for deployment, achieving significantly reduced inference latency and memory footprint with negligible accuracy degradation.
%
Table \ref{tab: quantization parameters} summarizes the parameter sizes of each module before and after edge optimization.
The combination of these optimization strategies can reduce the overall computational and memory footprint by more than 82.51\% compared to the unoptimized baseline, making practical deployment on commodity smartphones feasible.
All subsequent system tests are conducted on the optimized model on the end-side.


\begin{table}[htbp]
  \centering
  \caption{Module parameter sizes before and after edge optimization.}
  \scalebox{0.95}{
  \begin{tabular}{cccc}
    \toprule
    Module & Raw Model (MB)  & Edge Optimization (MB)\\
    \midrule
    Face Detection & 16.14  & 2.50 \\
    Face Encoder ($E_v$) & 166.31  & 13.61 \\
    Ultrasound Encoder ($E_u$) & 39.97  & 20.01 \\
    Fusion Module & 5.09  & 2.49 \\
    Watermark Embedder & 14.98  & 3.80 \\
    \textbf{Total} & \textbf{242.49} & \textbf{42.41} \\
    \bottomrule
  \end{tabular}
  }
\label{tab: quantization parameters}
\vspace{-0.1in}
\end{table}
\vspace{-0.1in}
\section{Dataset and Experimental Setup}
\subsection{Dataset Setup}\label{sec: dataset}

\begin{figure}[t]
  \centering
  \includegraphics[width=0.6\linewidth]{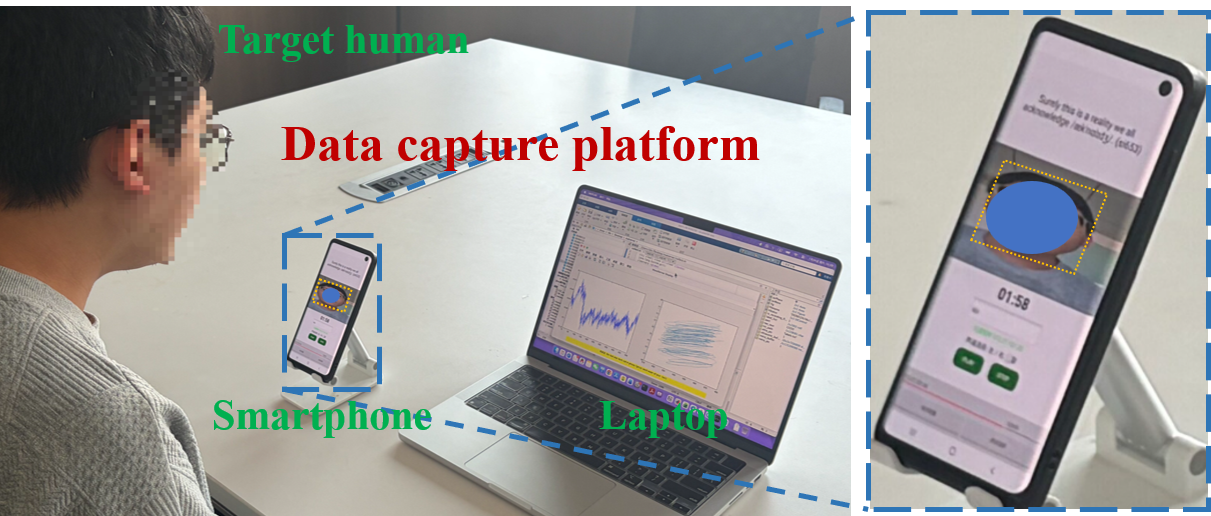}
  \caption{Data capture system and experimental setup.}
  \label{fig: people show}
   \vspace{-0.2in}
\end{figure}

We develop a synchronized video--ultrasound data acquisition system using MATLAB and Android Studio, deployed on a laptop and smartphones, respectively, as shown in Fig. \ref{fig: people show}. Using this system, we conduct extensive testing and data collection on multiple smartphone platforms, including Samsung S10 (S10), Samsung S22 Ultra (S22 Ultra), Vivo X200 Pro (Vivo), and OnePlus 15 (OnePlus).
Following Section \ref{sec: ultrasound preliminary}, we transmit 16 ultrasonic frequencies and convert the received echo at each frequency into In-phase (I) and Quadrature (Q) channels. The 32-channel IQ signals (16 frequencies $\times$ 2 channels) are concatenated to form the ultrasonic sample for a given time window $T$. The video is segmented along the same timeline using the same $T$ and paired with the corresponding ultrasonic sample to form a sample pair.
We recruit 50 volunteers (25 male, 25 female) and collect synchronized facial video--ultrasound pairs as legitimate samples. We randomly select 10 users for testing, with each user contributing 25--30 minutes of data, totaling approximately 25 hours. 
In the main experiments, we set $T$ to 5 seconds. During data collection, we maintain the distance between the user's face and the phone screen plane at 20--25 cm.
All experiments received our institutional review board (IRB) ethical approval before the start of the study.

\subsection{Attack Scenarios}\label{sec: Attack Scenarios}

Based on the threat model, we classify adversarial behaviors into two categories:

\subsubsection{Source-level attacks} 
These attacks attempt to forge authenticity evidence during video capture. We consider four representative attack scenarios:
\noindent \textbf{S1: Simple Video Replay.}
The target video is replayed without any corresponding ultrasonic signal input, simulating an attacker playing a pre-recorded video.
\noindent \textbf{S2: Random Pairing.}
The target video is randomly paired with ultrasonic samples from non-target users, simulating incorrect video--ultrasound pairing.
\noindent \textbf{S3: Advanced Video Replay.}
The target user's facial video is displayed on another device screen (e.g., a computer monitor) while synchronized ultrasonic signals are captured through the microphone, simulating a replay attack with the attacker's device.
\noindent \textbf{S4: Sophisticated Video Forgery.}
We employ an existing face-swapping model \cite{wang2025dynamicface} to replace the face in a legitimate sample video while using the synchronized ultrasonic signal from the attacker as input, to simulate a real-time face-swapping attack.

For these negative samples, we construct datasets of 1 hour, 125 hours, 0.5 hours, and 1 hour, respectively. 

\subsubsection{Propagation-level attacks} 
Besides, other attacks manipulate video content after generation and transmission. We consider two representative propagation-level attack scenarios:
\noindent \textbf{P1: Forgery (Watermark Reuse).}
The adversary obtains a target user's legitimate facial videos and corresponding watermark information (accessible via our public detector) from the internet. They create a forged video of the target human face, inject the obtained watermark into it, and attempt to deceive other users. We generate such attack samples for different users, totaling approximately 7 hours.
\noindent \textbf{P2: Editing.}
The adversary performs localized edits or modifications to the face within a video while keeping the watermark information unchanged. We utilize the Edit-Yourself algorithm \cite{flynn2026edityourself} to edit collected facial videos, altering the user's mouth movements to convey a meaning completely different from the original, while preserving the original watermark. We create such edited samples for all users, totaling approximately 0.5 hours. Due to the powerful capability of face-swapping models, it becomes difficult for the human eye to distinguish between real and fake faces after swapping.
Including all attack samples from different users, the total duration exceeds 135 hours.

\vspace{-0.1in}
\subsection{Training Details}

For all experiments, model training is conducted on an NVIDIA RTX 4090 GPU.
%
We train the ultrasonic encoder $E_u$ with both the similarity loss and the MSE loss, where $\lambda_c$ and $\lambda_m$ are both set to 1. The learning rate is 5e-4, epochs 400, batch size 512, using AdamW with a CosineAnnealingLR scheduler triggered every 10 epochs. During this stage, $E_v$ remains frozen, and only $E_u$ and $D_u$ are updated.
We train the fused feature representation with the triplet loss, where the margin $\alpha$ is set to 1.0. The learning rate is 5e-4, batch size is 256. During this stage, $E_v$ and $E_u$ remain frozen, and only the fusion-related modules are trained.
For the watermark--video fusion module and the watermarking detector, we follow the training protocol of InvisMark and fine-tune its pre-trained open-source model on our task and dataset.
%
%
To enable efficient edge deployment on mobile platforms, we perform QAT starting from the pre-trained FP32 model for the INT8 module. Specifically, we adopt the AdamW optimizer with an initial learning rate of 1e-5, a weight decay of 1e-5, and cosine-annealed learning-rate scheduling. The model is fine-tuned for 200 epochs with a batch size of 16, where simulated quantization operations are inserted during training to approximate INT8 inference behavior. The QNNPACK backend is used as the quantization engine for mobile deployment.

\section{Result and Evaluation}
\subsection{Evaluation Metrics}

We extensively evaluate our system's performance using standard metrics: True Positive Rate (TPR), False Positive Rate (FPR), True Negative Rate (TNR), and False Negative Rate (FNR). We also employ the Equal Error Rate (EER), representing the operating point where FPR equals FNR, and the Area Under the ROC Curve (AUC), along with accuracy, defined as the proportion of correctly classified instances within the corresponding test category. The Hamming distance \cite{hamming1950error} is used for watermark similarity comparison in propagation-level attack scenarios.

\begin{figure}[ht]
\begin{minipage}[t]{0.36\linewidth}
    \centering
    \includegraphics[width=\linewidth]{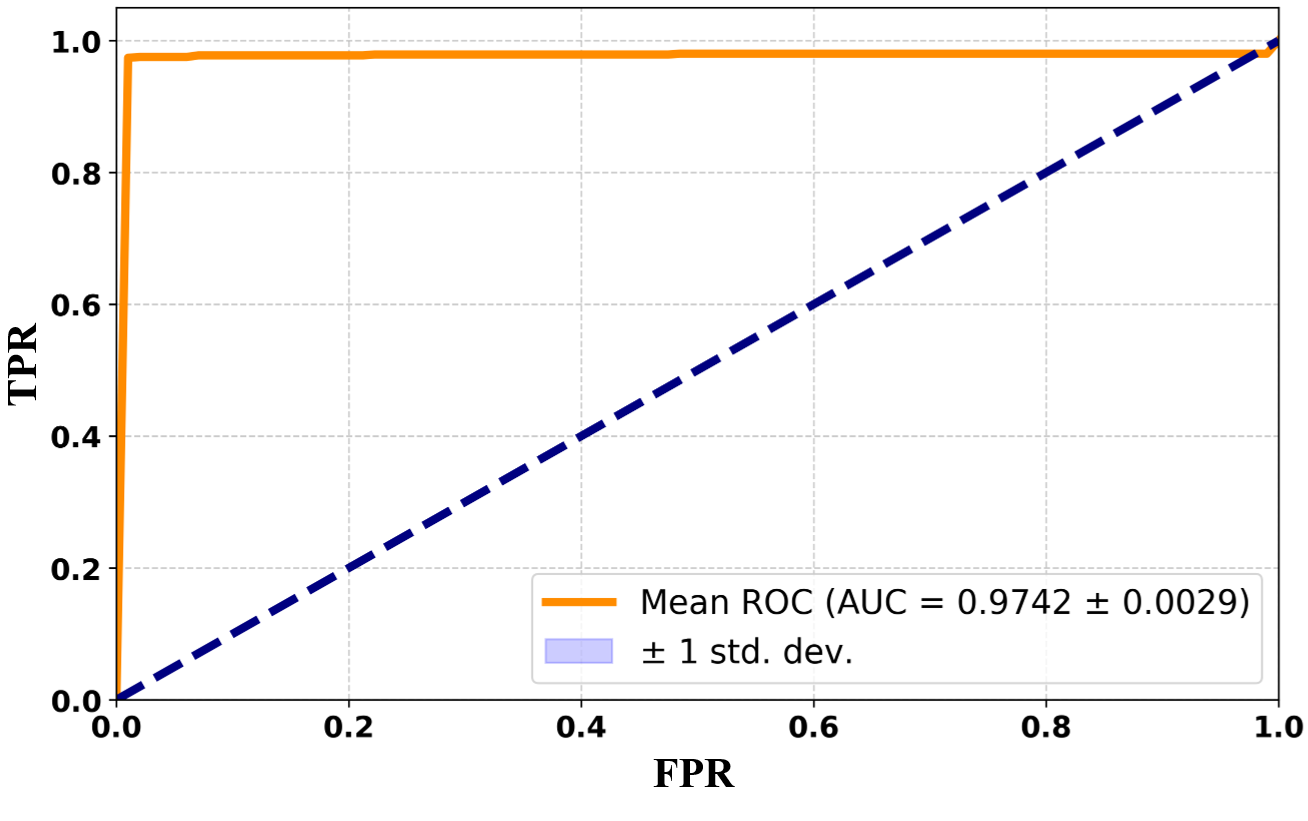}
    \caption{The ROC curve and AUC result.}
    \label{fig: result_roc}
\end{minipage}
\hspace{0.01\linewidth}
\begin{minipage}[t]{0.39\linewidth}
    \centering
    \includegraphics[width=\linewidth]{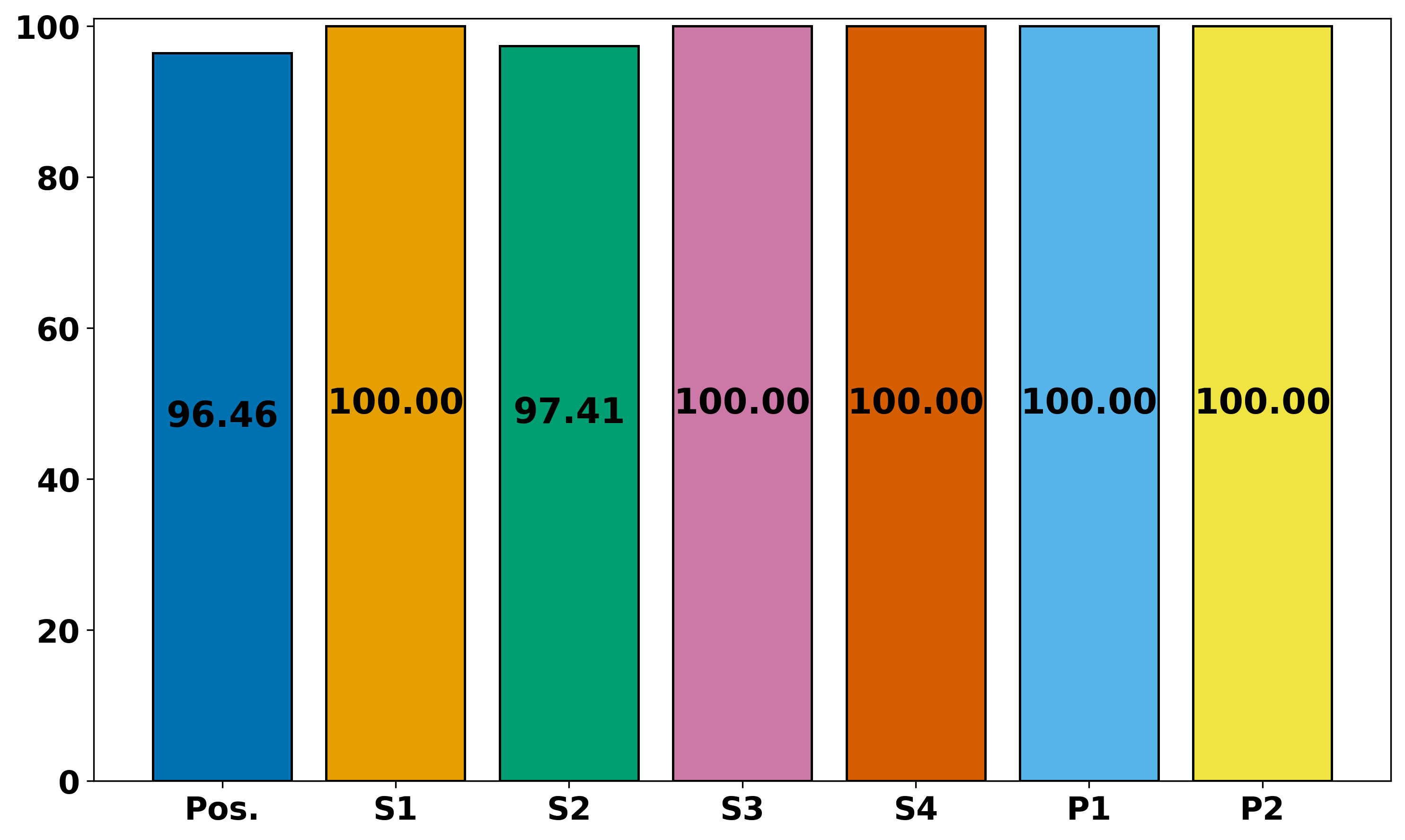}
    \caption{The accuracy of different samples.}
    \label{fig: result_bar}
\end{minipage}
\vspace{-0.2in}
\end{figure}

\subsection{Overall Performance}
To evaluate the accuracy for legitimate and different attack samples, we randomly select samples from test users as enrollment samples and use the remaining for comparison. 
We randomly select enrollment samples 20 times, iteratively compute the average ROC curve, and derive the mean AUC and EER values. The results are shown in Fig. \ref{fig: result_roc}, yielding a mean AUC of 0.9742 $\pm$ 0.0029 and a mean EER of 0.0225. 
The accuracy results for legitimate samples are presented on the left of Fig. \ref{fig: result_bar}, achieving 96.46\% accuracy.

After model edge optimization, we apply INT8 quantization with QAT to the watermark embedding module for efficient on-device deployment, while keeping the watermark detection module unchanged since it runs on the server side. We perform 100 independent watermark embedding and detection trials to evaluate the impact of QAT. The detection accuracy improves from 87.01 ± 6.31\% to 95.48 ± 2.12\% after QAT, demonstrating that QAT effectively mitigates quantization-induced degradation and enables reliable on-device watermark embedding.

%
%

\subsection{Source-level Attack Evaluation}

 For different source-level attack sample scenarios, maintaining an average 99.35\% accuracy, as shown in the middle of Fig. \ref{fig: result_bar}. This demonstrates the effectiveness of our system's source-level protection. With such high accuracy, our defense mechanism can effectively prevent attackers from forging content at the source, as an attacker must simultaneously satisfy video, ultrasonic, and fusion feature similarity thresholds---a combinatorial constraint that is extremely difficult to meet without genuine biometric capture.

The high accuracy against attack scenarios S1 and S2 benefits from our system's cross-modal consistency constraints: replay attacks lack corresponding synchronized ultrasound, and random pairing explicitly violates the temporal alignment between modalities. For S3, although the video is displayed on another device, the captured ultrasonic reflections still differ from genuine face reflections in two critical ways: first, the flat screen surface produces specular rather than diffuse acoustic reflections; second, the material properties (glass vs. skin) result in different acoustic impedance and frequency-dependent attenuation patterns. Our multi-frequency CW design (16 sub-carriers) is particularly effective at detecting these discrepancies because different frequencies probe different surface depths and geometries. For S4, face-swapping alters the facial content while the ultrasound remains tied to the original face's biometric structure, creating a detectable cross-modal mismatch.


\subsection{Propagation-level Attack Evaluation}

In propagation-level attacks, adversaries primarily target the watermark information within the video. By converting the detected video into a corresponding dynamic watermark and comparing it with the embedded watermark via binary encoding, video authentication can be performed. A watermark similarity measure based on the Hamming distance is applied.
For P1, since the facial information is entirely fabricated, the resulting watermark encoding differs significantly from the correct one, making it easily identifiable. The overall test accuracy remains at 100\%, indicating these forgeries are readily detected.

For P2, when submitted to our cloud-based detection system for verification of the content-dependent dynamic Face Key, our system readily identifies the mismatch between the tampered video's encoding and the original dynamic watermark. The overall average similarity score is only 82.33 ± 5.31\%, with no samples exhibiting identical watermark information. Consequently, the negative sample detection accuracy achieves 100\%. This is because our Face Key encoding is derived from a holistic representation of facial content within the time window $T$---any localized modification to facial features, such as altered mouth movements, produces a different feature vector that maps to a different Face Key encoding. The fixed User Key would remain unchanged in such attacks, but the dynamic Face Key provides the crucial second layer of protection that catches these sophisticated edits. This demonstrates that our system and the proposed spatio-temporal dynamic encoding watermark can effectively resist these meticulously crafted attacks, ensuring robust security for users' facial videos even against state-of-the-art editing tools.


\begin{figure}[t]
    \centering
    \begin{minipage}[t]{0.30\linewidth}
        \centering
        \includegraphics[width=\linewidth]{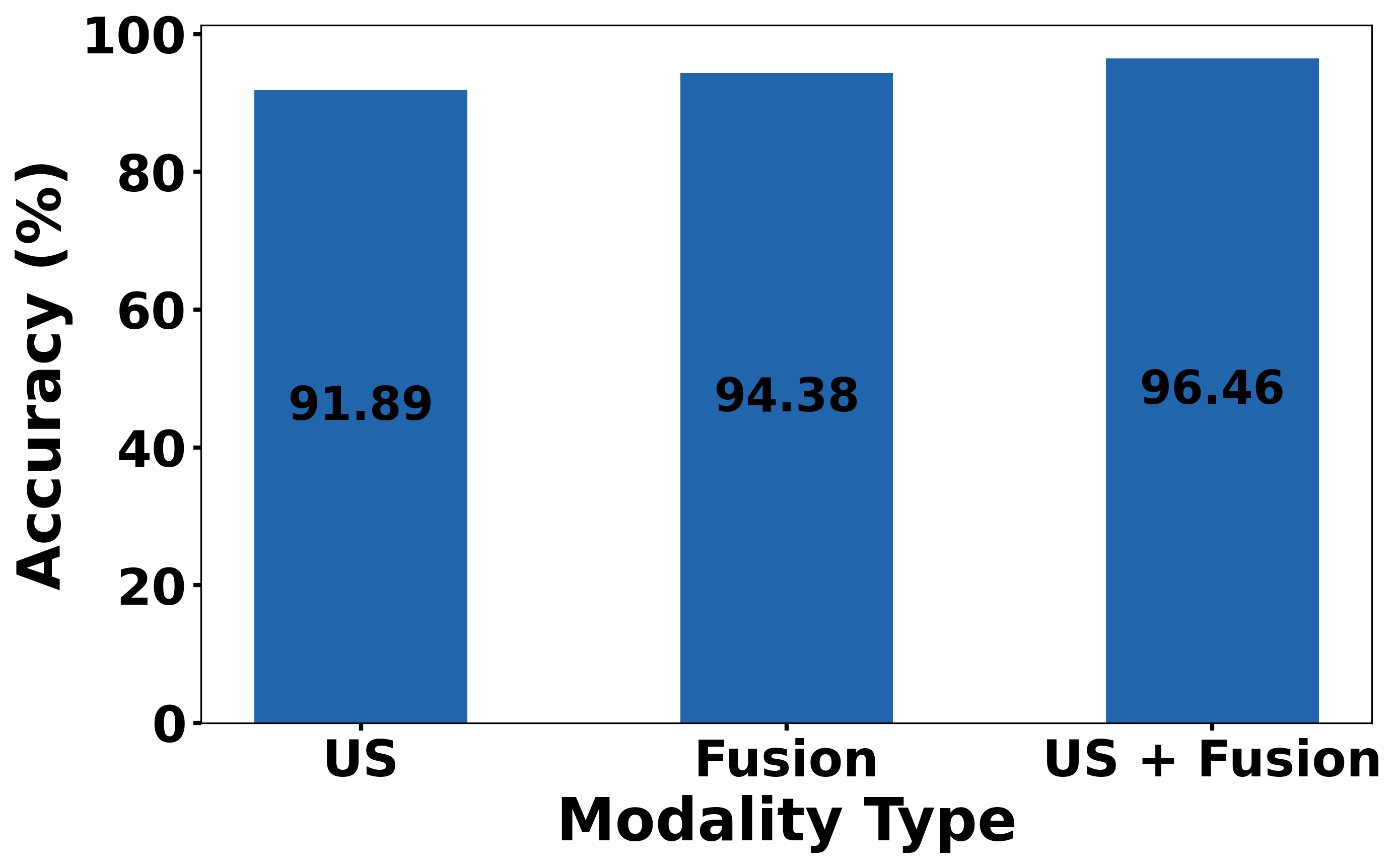}
        \caption{Different modality types.}
        \label{fig: Impact modality}
    \end{minipage}
    \hspace{0.02\linewidth}
    \begin{minipage}[t]{0.30\linewidth}
        \centering
        \includegraphics[width=\linewidth]{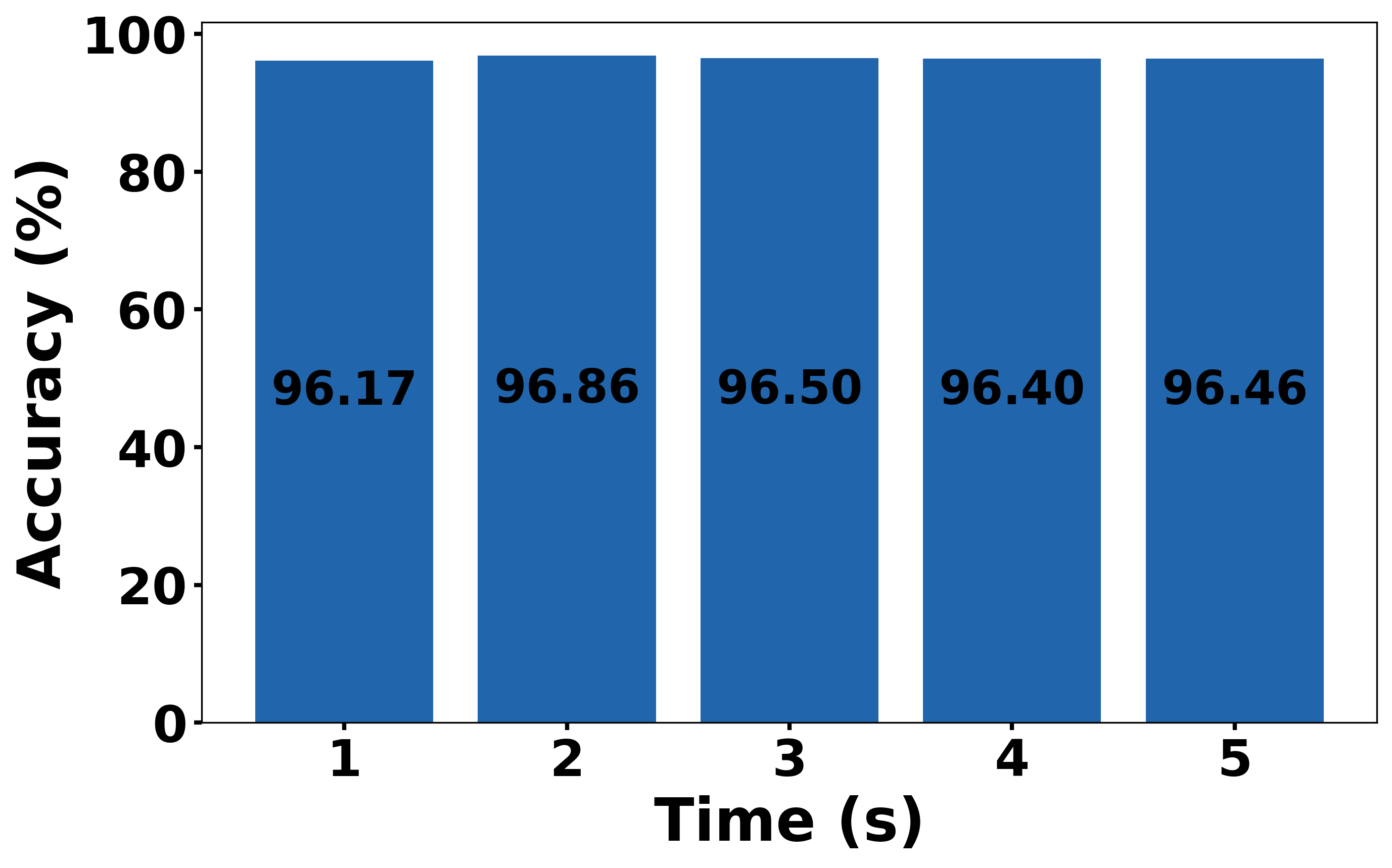}
        \caption{Different time windows.}
        \label{fig: Impact time}
    \end{minipage}
    \hspace{0.02\linewidth}
    \begin{minipage}[t]{0.30\linewidth}
        \centering
        \includegraphics[width=\linewidth]{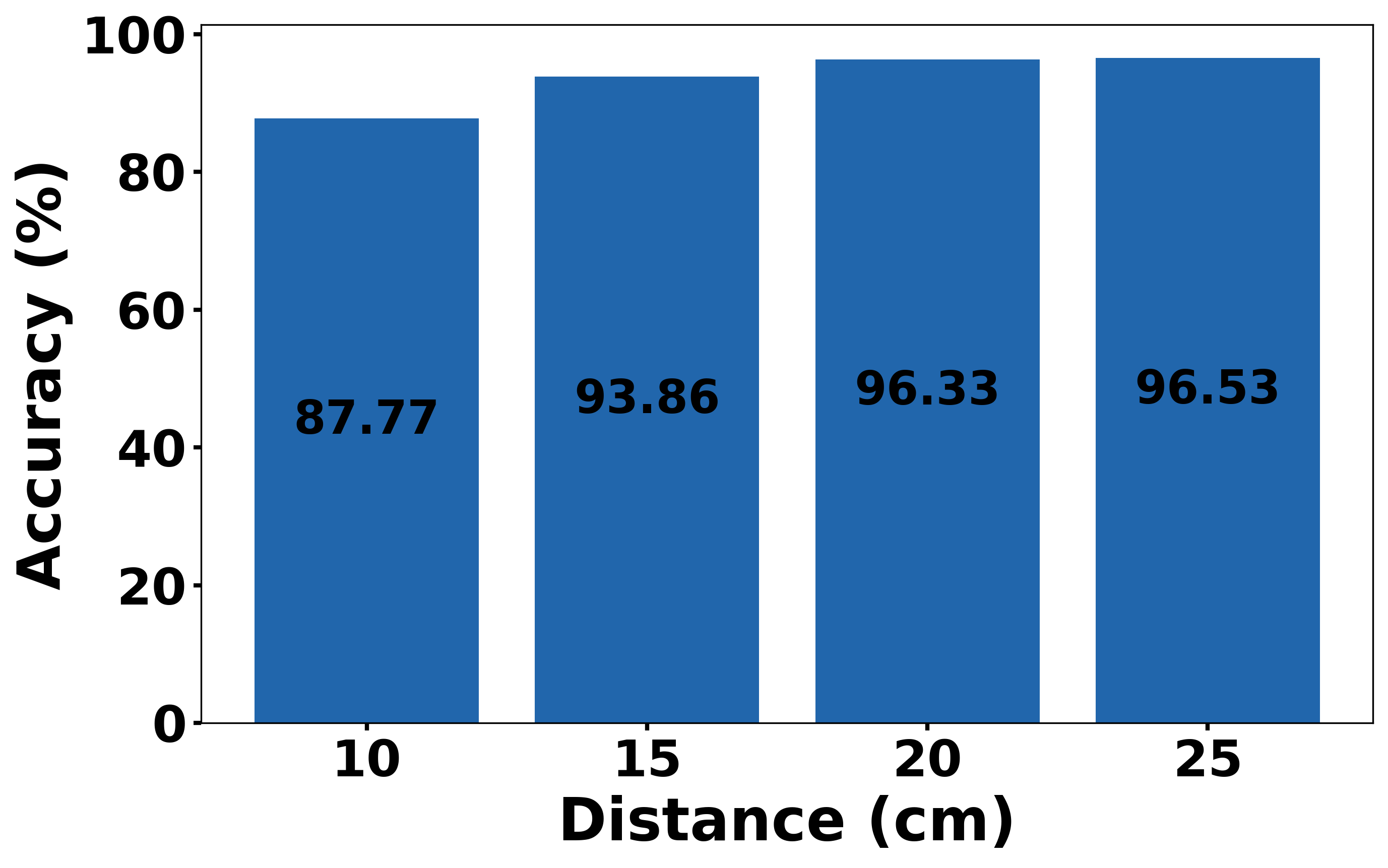}
        \caption{Different distances.}
        \label{fig: Impact distance}
    \end{minipage}
    \vspace{-0.2in}
\end{figure}

\begin{figure}[t]
    \centering
    \begin{minipage}[t]{0.30\linewidth}
        \centering
        \includegraphics[width=\linewidth]{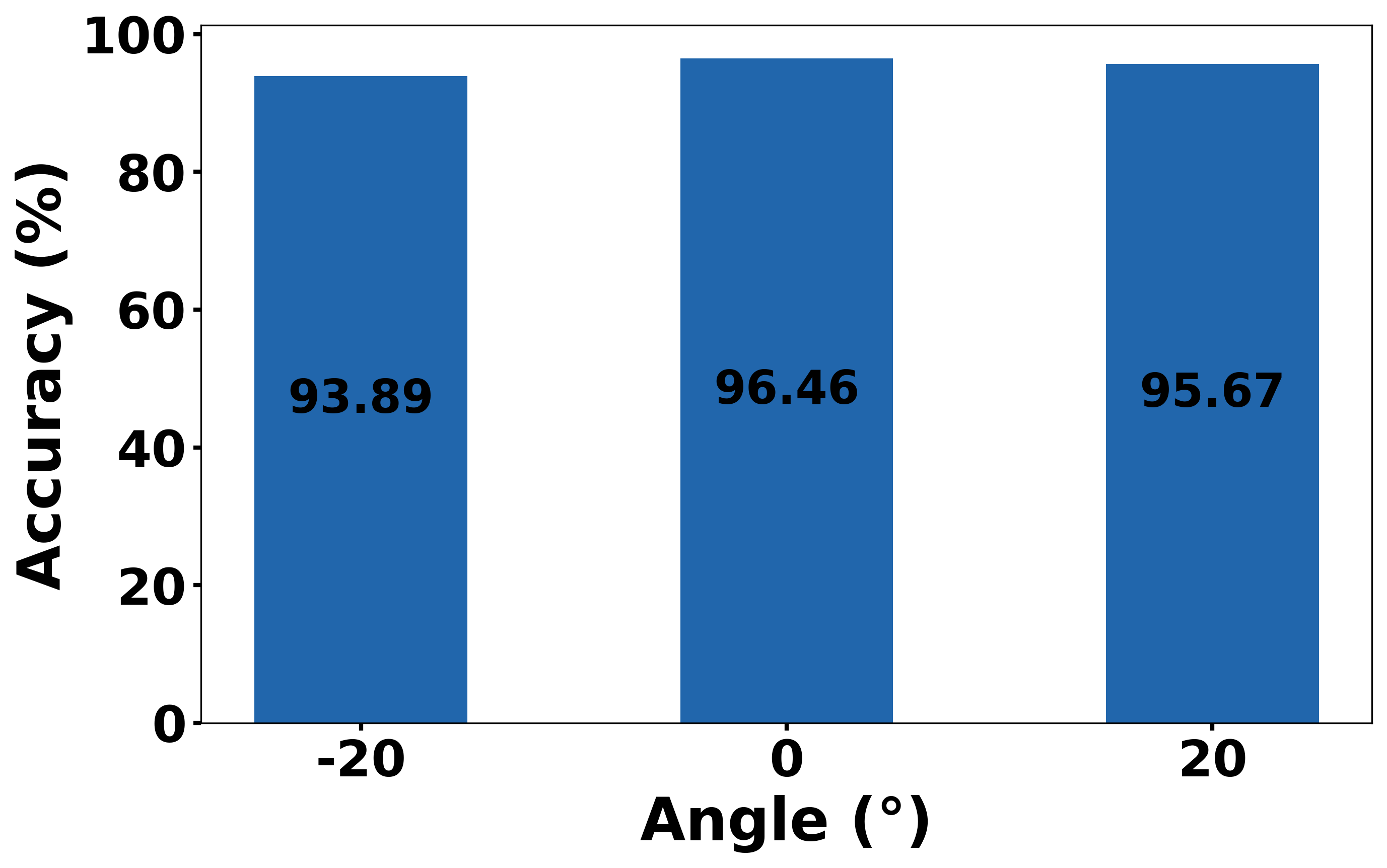}
        \caption{Different angles.}
        \label{fig: Impact angle}
    \end{minipage}
    \hspace{0.04\linewidth}
    \begin{minipage}[t]{0.30\linewidth}
        \centering
        \includegraphics[width=\linewidth]{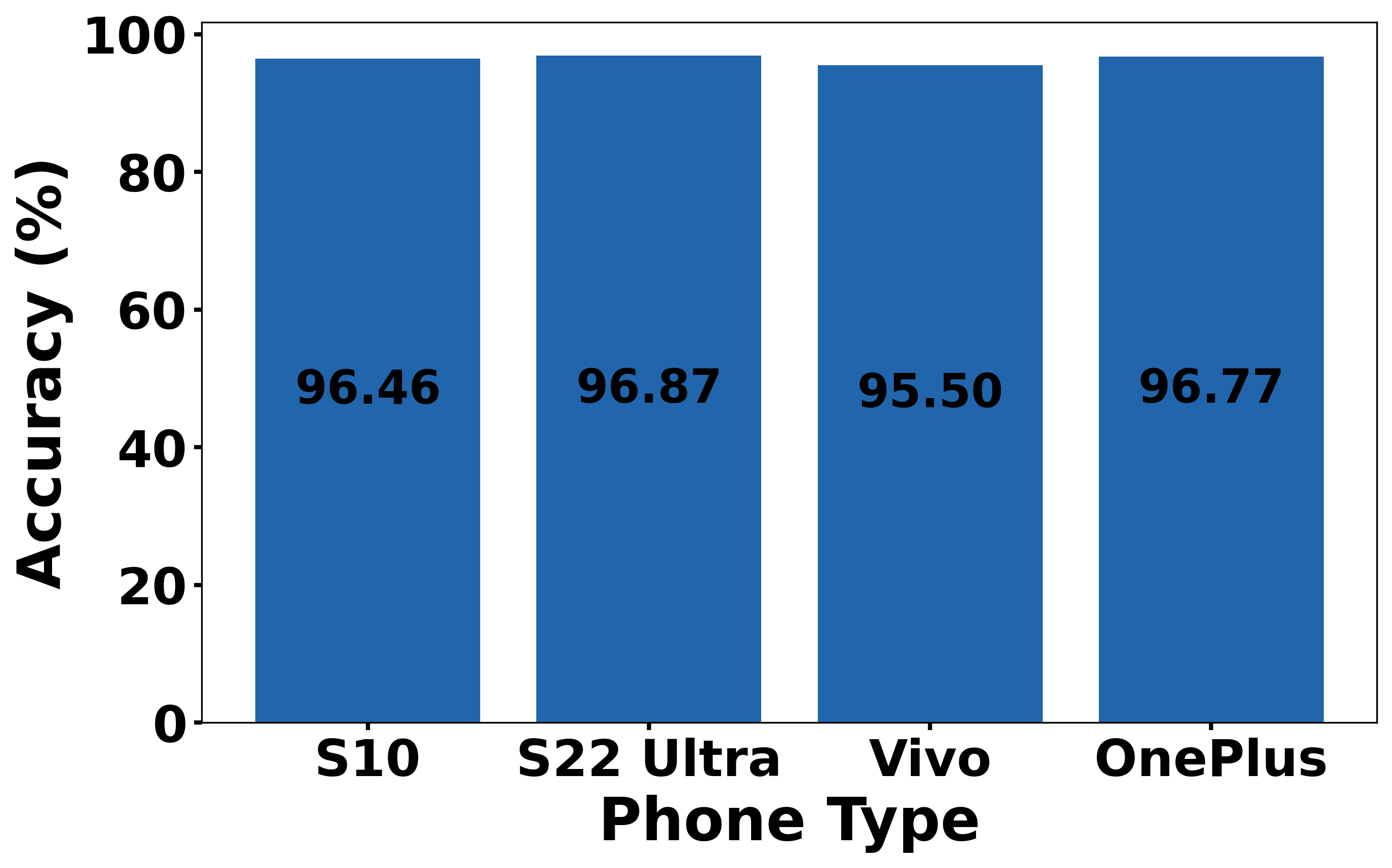}
        \caption{Different phone types.}
        \label{fig: Impact phone}
    \end{minipage}
\end{figure}

\subsection{Evaluation on Impact Factors}
For the various influencing factors considered in this experiment, these factors are only related to the discrimination of source input, so we use legitimate sample accuracy to evaluate them.
It is worth noting that the distance and angle between the face and the smartphone need to be within the effective propagation range of ultrasound. This range can cover the usage distance in normal video call scenarios, i.e., the camera can fully capture the user's face region and ensure that it occupies a sufficient proportion in the image. Unlike purely software-defined watermarking schemes, the coverage of ultrasound is constrained by the physical laws of sound wave propagation. Therefore, successful watermark embedding depends not only on the video and ultrasound signals themselves but also on the active participation of the user and the satisfaction of an effective spatial geometric relationship between the device and the face. From this perspective, such a physical constraint is not a system limitation but rather an additional physical security mechanism.
In authentication systems, tighter decision boundaries generally lead to stronger robustness against attacks, since the consequence of falsely accepting an attacker is considerably more severe than that of falsely rejecting a legitimate user.

\subsubsection{Impact of Different Modal Features}
In the decision logic for verifying the genuine target, we utilize a joint judgment based on both the ultrasound (US) and fusion features. The results in Fig. \ref{fig: Impact modality} show that the highest accuracy is achieved when both features are combined (96.46\%), compared to video-only (94.20\%), ultrasound-only (89.14\%), or simple concatenation (93.87\%). This indicates that the trained fusion feature effectively learns the correspondence between video and ultrasonic modalities, rather than merely performing a simple addition. Furthermore, the video modality does not dominate or suppress the ultrasonic modality, a risk inherent in multimodal systems with imbalanced information content, highlighting the advantage of our late-fusion design.

\subsubsection{Impact of Different Time Windows}\label{sec: different time window}
The time window length $T$ is a key factor in our dynamic video watermarking system. It affects both the accuracy of ultrasonic discrimination and the cost of watermark generation. A larger $T$ yields a longer watermark constraint window and reduces the computation overhead because the watermark updates less frequently. In contrast, a smaller $T$ lowers latency but increases the generation cost and may affect discrimination accuracy. As shown in Fig. \ref{fig: Impact time}, the system remains stable across different window lengths, with an average accuracy of 96.48\%. Even when $T=1$ s, the system still achieves 96.17\% accuracy. These results show that our system can adapt $T$ to balance temporal dynamic encoding, recognition accuracy, and latency in practice.

\subsubsection{Impact of Different Distances}\label{sec: different distance}
During video calls, users hold their phones at varying distances. Due to hardware limitations of smartphones and the inherent attenuation properties of ultrasonic signals, the received echo strength is influenced by these distance variations. We evaluate system performance at 10 cm, 15 cm, 20 cm, and 25 cm, which comprehensively cover the typical range of distances users maintain during normal video calls. The test results in Fig. \ref{fig: Impact distance} show effective robustness, with overall accuracy consistently above 87.77\% and an average of 93.62\%. The accuracy at 10 cm drops to 87.77\% primarily because the camera captures only a partial view of the face at such close range, reducing the effectiveness of video feature extraction. At 15 cm and beyond, the accuracy stabilizes above 93\%, indicating that the system performs robustly at typical video call distances.

\subsubsection{Impact of Different Angles}\label{sec: different angle}
The angle formed between the smartphone screen plane and the user's head is another metric we evaluate. In real-world video call scenarios, users often hold their phones at various angles relative to their faces. We define the parallel state between the phone screen plane and the facial plane as 0 degrees. We then rotate the phone upward and downward to create positive and negative angle differences, collecting data within approximately $\pm$20 degrees. This range ensures that the phone camera can effectively capture the user's face while the ultrasonic signals can be properly transmitted and received. The results in Fig. \ref{fig: Impact angle} show accuracy consistently above 93.89\%, demonstrating that moderate angular variations do not significantly impact system performance. This robustness benefits from the wide acoustic sensing field of the top earpiece-microphone pair, which maintains sufficient overlap with the facial region across the tested angle range.

\subsubsection{Impact of Different Smartphone Types}
We evaluate across four smartphone models spanning different manufacturers and hardware generations: Samsung S10 (S10), Samsung S22 Ultra (S22 Ultra), Vivo X200 Pro (Vivo), and OnePlus 15 (OnePlus). Test results in Fig. \ref{fig: Impact phone} show effective verification across all devices, with accuracy consistently above 95.50\% and an average of 96.42\%. This robustness stems from the fact that ultrasonic signals rely only on the microphone and earpiece speaker, both of which are standard, mature components present in all smartphones regardless of price tier. The relatively small variance across devices confirms that UltraGuard is broadly deployable without per-device calibration or tuning.

\subsection{System Latency and Privacy Security}

\begin{figure}[ht]
\begin{minipage}[t]{0.40\linewidth} 
    \centering
    \includegraphics[width=\linewidth]{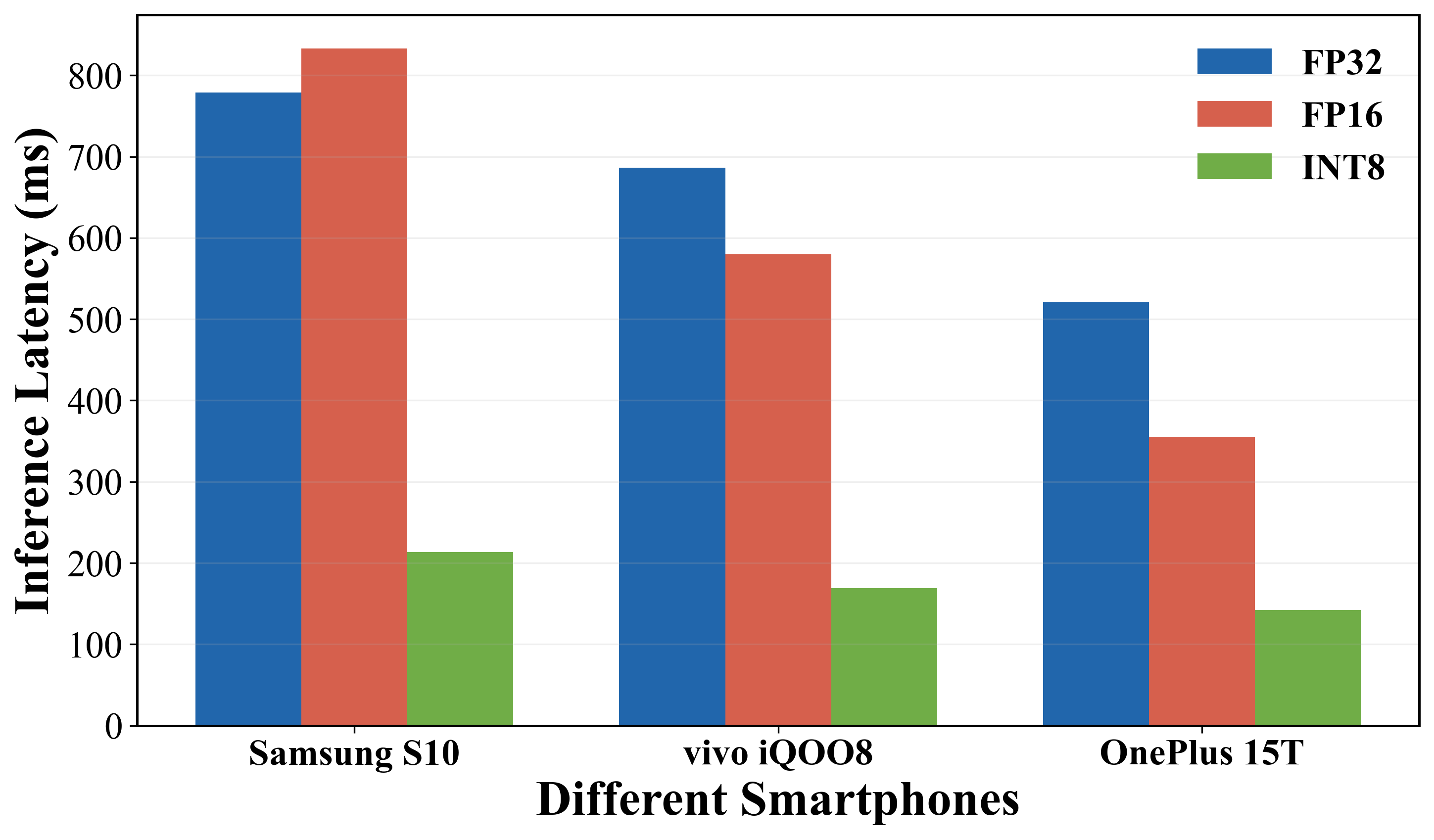}
    \caption{Inference latency.}
    \label{fig: inference latency}
\end{minipage}
\hspace{0.01\linewidth}
\begin{minipage}[t]{0.40\linewidth} 
    \centering
    \includegraphics[width=\linewidth]{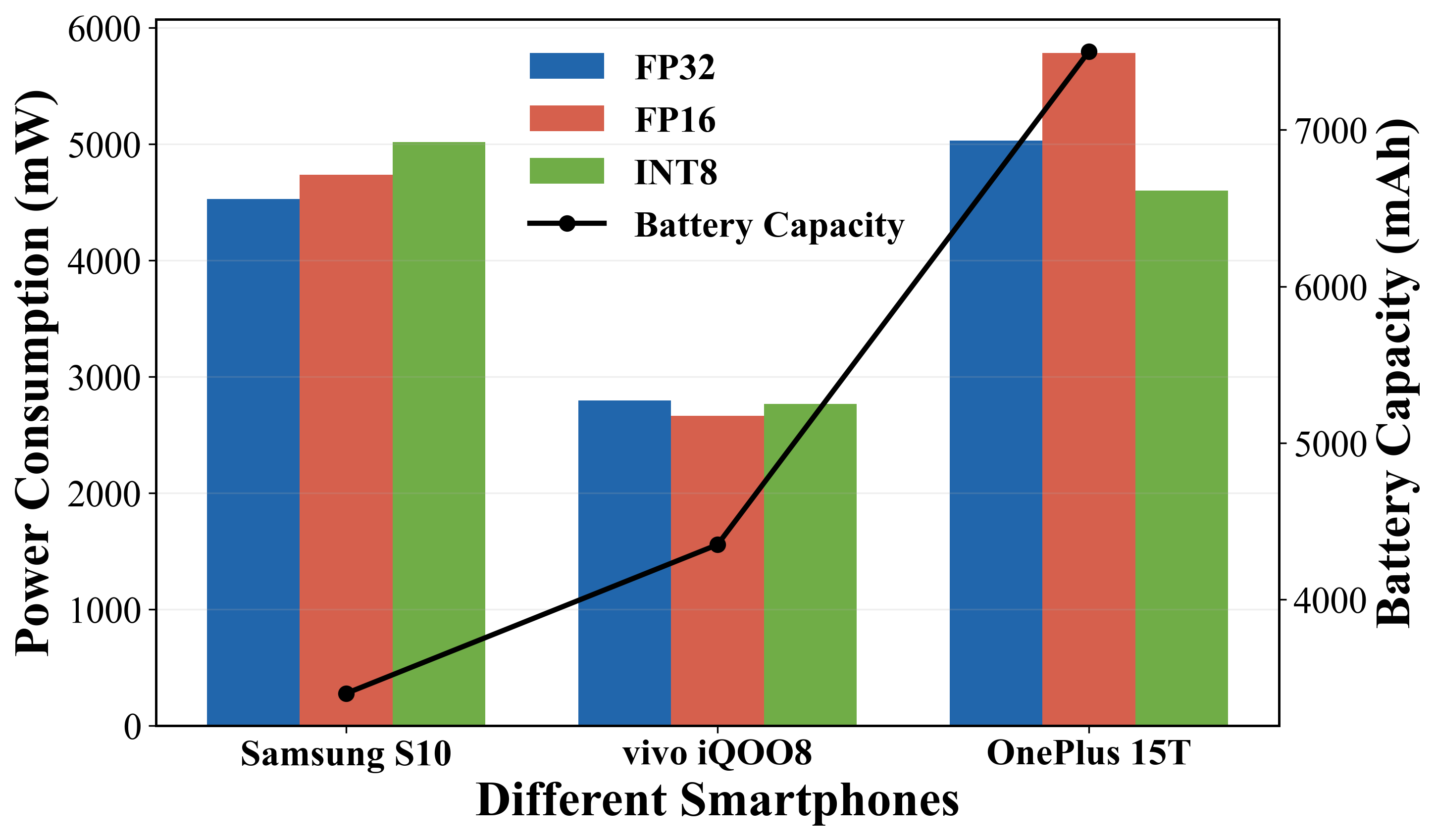}
    \caption{Power consumption.}
    \label{fig: power consumption}
\end{minipage}
\vspace{-0.2in}
\end{figure}

We deploy UltraGuard on three representative smartphones, namely the Samsung S10 (2019), vivo iQOO8 (2021), and OnePlus 15T (2026), and evaluate its latency and power consumption under simulated real-time video call scenarios. 
Using the optimization scheme for on-device video watermarking described in Section \ref{sec: edge deployment}, for a 30-frame-per-second video, we set the number of key frames in the cache to K=5, use the average feature of these frames to replace the features of all frames within the 1-second window, and embed the watermark only into the most critical facial lips region.
Average results obtained from 100 test samples under different quantization settings are summarized in Fig.~\ref{fig: inference latency} and Fig.~\ref{fig: power consumption}.
All experiments are conducted on CPUs to ensure broad deployability. Even without GPU or NPU acceleration, UltraGuard achieves a minimum end-to-end latency of 213.64 ms, 168.91 ms, and 142.40 ms on three smartphones, respectively.
This means that within a duration of 1 second, watermarks can be applied to over 5 key frames, which covers a real-time processing speed, thereby meeting the requirements~\cite{series2003transmission} of real-time video call processing workflows.
%
Moreover, with an average power consumption, these devices can operate continuously for 2.89, 6.29, and 6.28 hours, respectively.
These results demonstrate the feasibility of practical real-time deployment on commodity edge devices for video watermarking.

From a security perspective, the watermarking pipeline is executed primarily on the device, while cloud interaction is limited to user enrollment and identifier assignment. User biometrics are represented as irreversible embedding vectors before transmission, thereby mitigating privacy risks. For cloud-assisted dynamic watermark verification, standard hybrid encryption schemes~\cite{ietf_tls13_2018, nist_aes_2001, rivest_rsa_1978} can be employed to ensure secure communication.
%

\section{Discussion}

As AIGC technologies continue to advance, protecting the authenticity and integrity of real-time facial video becomes increasingly important. UltraGuard explores ultrasound-assisted biometric watermarking as a promising approach to this challenge. Nevertheless, several limitations remain and provide opportunities for future improvement.

First, UltraGuard relies on ultrasonic echoes to capture biometric features, which limits the effective sensing distance. As the user moves farther from the smartphone, weaker echoes may degrade feature quality and reduce authentication reliability. In addition, substantial head movements and frequent facial occlusions may disrupt the temporal consistency between ultrasonic signals and video features, leading to authentication failures. Although these constraints limit operational flexibility, the physical interaction required for ultrasonic sensing also introduces an additional barrier to impersonation. In security-sensitive scenarios, conservative authentication thresholds can help reduce false acceptance at the cost of potentially rejecting some legitimate samples. Future work could incorporate complementary sensing modalities to improve robustness under dynamic conditions while preserving reliable authentication.

Second, the security of the sensing pipeline depends on the integrity of the underlying hardware and data acquisition process. Although UltraGuard provides protection through ultrasound-assisted authentication and video watermarking, unauthorized access to internal sensor data or the injection of forged data into the acquisition pipeline remain potential concerns if the hardware is compromised. Future work could integrate hardware-assisted encryption and data-flow integrity verification to strengthen protection against such attacks.

Third, video compression during transmission may degrade watermark integrity and impair subsequent verification. UltraGuard mitigates this challenge by leveraging adversarial training and feature-level fusion to embed watermarks into content-correlated representations, providing a degree of resilience to compression \cite{fernandez2024video}. Further improvements could be achieved by integrating watermark embedding more closely with the video encoding pipeline, enabling the watermarking process to account for codec-induced distortions more directly.

Fourth, the current dataset does not fully represent the diversity of users, demographic groups, and environmental conditions encountered in real-world deployments. Moreover, the evaluated attack scenarios do not cover all possible threats. Future work should expand dataset diversity and investigate a broader range of attack strategies to better assess and improve the system's generalizability and robustness.

Despite these limitations, UltraGuard provides a foundation for exploring the integration of physical sensing and biometric watermarking for real-time video protection. To the best of our knowledge, it is the first approach to combine ultrasound and video for real-time biometric video watermarking. The underlying framework may also be extended to other sensing modalities and video content beyond facial videos, offering opportunities for broader protection against the misuse of AIGC-generated content.

\section{Conclusion}

We propose UltraGuard, a novel facial video watermarking scheme that leverages a smartphone’s built-in speaker and microphones to emit ultrasonic signals and capture biometric features, which are then fused with facial video content to generate content-correlated dynamic watermarks without perceptible visual degradation.
Designed for real-time video communication, UltraGuard establishes source-level system modeling and content-dependent dynamic encoding to achieve end-to-end security throughout video generation and propagation.
Extensive experiments validate its effectiveness, and it can be widely deployed on mobile devices to defend against AIGC and other video forgery threats to facial privacy and security.

\bibliographystyle{ACM-Reference-Format}
\bibliography{sample-base}

\appendix

\end{document}